\documentclass[twocolumn,showpacs,amsmath,amssymb]{revtex4}

\usepackage{graphicx}
\usepackage{dcolumn}
\usepackage{bm}

\begin{document}

\preprint{APS/123-QED}

\title{Phase transition of the three-dimensional chiral Ginzburg-Landau model 
\\  --- search for the chiral phase}

\author{Tsuyoshi Okubo}
 \email{okubo@spin.ess.sci.osaka-u.ac.jp}
\author{Hikaru Kawamura}%
\affiliation{Department of Earth and Space Science,
 Faculty of Science, Osaka University, Toyonaka, Osaka 560-0043%
}%
\date{\today}

\begin{abstract}
 Nature of the phase transition of regularly frustrated vector spin systems in three dimensions is investigated based on a Ginzburg-Landau-type effective Hamiltonian.  On the basis of the variational analysis of this model, Onoda {\it et al} recently suggested the possible occurrence of a {\it chiral phase}, where the vector chirality exhibits a long-range order without the long-range order of the spin [Phys. Rev. Lett. {\bf 99},  027206  (2007)]. In the present paper, we  elaborate their analysis by considering the possibility of a first-order transition which was not taken into account in their analysis. We find that the first-order transition indeed occurs within the variational approximation, which significantly reduces the stability range of the chiral phase, while the chiral phase still persists in a restricted parameter range. Then, we perform an extensive Monte Carlo simulation focusing on such a parameter range. Contrary to the variational result, however, we do not find any evidence of the chiral phase. The range of the chiral phase, if any, is estimated to be less than $0.1\%$ in the temperature width.
\end{abstract}

\pacs{75.10.Hk, 05.50.+q, 75.40.Mg, 64.60.F-}
\maketitle

\section{\label{sec:Intro} Introduction}

In vector spin systems, frustrations often induces noncollinear or noncoplanar spin structures. Such canted spin structures generally accompany the order of the chirality \cite{Villan}. Two types of chirality has been discussed in the literature. One is a scalar chirality which is defined as a scalar product of three Heisenberg spins, $\chi \sim \vec{S}_i\cdot (\vec{S}_j\times\vec{S}_k)$. The scalar chirality takes a nonzero value for a noncoplanar spin configuration. By contrast, a vector chirality, which is a target of this paper, is defined as a vector product of two Heisenberg (or {\it XY\/}) spins, $\vec{\kappa} \sim \vec{S}_i \times \vec{S}_j$. It takes a nonzero value even for a noncollinear but coplanar spin configuration. The ordering of the vector chirality is realized in, {\it e.g.\/}, conventional helical magnets.

 Although the chiral order inevitably appears in noncollinear or noncoplanar spin ordered states, it can be realized in principle without accompanying the long range order of the spin. In such chiral ordered but spin disordered state, spin correlation lengths are kept finite while the chirality shows a long-range order. 

 In the past, the existence of such a {\it chiral phase} has been
 discussed for several frustrated vector spin systems including spin
 glasses \cite{Kawamura_SG_rev,Kawamura, Hukushima, Viet} and regularly frustrated
 magnets \cite{Miyashita, Ozeki, Hasenbusch, Kawamura_XY,Onoda}. For example, it has been suggested  in the three-dimensional (3D) Heisenberg spin glass that the glass-order of the scalar chirality takes place at a temperature higher than that of the spin-glass order \cite{Kawamura, Hukushima, Viet}.
In regular systems, it has been suggested that the two-dimensional (2D) fully frustrated {\it XY\/} models exhibits the ordering of the vector chirality at a temperature higher than that of the spin Kosterlitz-Thouless transition \cite{Miyashita,Ozeki,Hasenbusch}. Such  occurrence of separate chiral and spin transitions is often called ``spin-chirality decoupling''.

 In regularly frustrated 3D systems, however, there has been no clear
 evidence of such an intermediate chiral phase so far. For classical
 Heisenberg and {\it XY\/} antiferromagnets on the 3D stacked-triangular
 lattice, Monte Carlo (MC) simulations suggested the occurrence of a
 single magnetic phase transition from a paramagnetic phase to a helical
 magnetic phase
 \cite{Kawamura_rev,Kawamura_1992,MC_stack2,MC_stack3,MC_first,MC_stack4}. Based
 on a renormalization-group (RG) analysis, Kawamura suggested that the
 phase transition of noncollinear magnets could belong to a new
 ``chiral'' universality class distinct from the standard $O(N)$
 Wilson-Fisher universality class, whereas the transition could also be
 of first-order depending on the parameter values of the system
 \cite{Kawamura_1988}. Some supports to this scenario were reported from
 field theoretical approaches \cite{Vicari1,Vicari2} and Monte Carlo
 (MC) simulations \cite{Kawamura_1992,MC_stack2,MC_stack3, Vicari2}. Experimental measurements of relevant critical exponents also seem consistent with such a chiral universality class \cite{Exp1,Exp2,Exp3,Exp4,Exp5}. By contrast, some authors argued that the transition might be weakly first-order \cite{stack_first,MC_first, MC_stack4}. In any case, though the nature of the transition has been somewhat controversial, it has been believed that an intermediate chiral phase does not appear in 3D regular systems.

 Recently, Onoda and Nagaosa studied the possibility of the vector chiral phase in regularly frustrated 3D Heisenberg systems \cite{Onoda}. Based on a Ginzburg-Landau (GL)  Hamiltonian describing helical Heisenberg magnets and performing  variational calculations, these authors suggested that the Dzyaloshinskii-Moria interaction and/or the Coulombic four-spin ring-exchange interaction could stabilize the chiral phase even in 3D.

In this paper, motivated by the recent work by Onoda {\it et al}, we wish to examine the nature of the phase transition of the same GL Hamiltonian as studied by Onoda  {\it et al} by means of a further analytical calculation and a MC simulation on a discretized version of the model, with particular attention to the issue of the existence/non-existence of an intermediate chiral phase. Note that Onoda {\it et al\/} implicit assumed in their analysis a continuous nature of the transition, ignoring the possibility of a first-order transition\cite{Onoda}. We see in the present paper that a first-order transition indeed occurs within the variational approximation, which significantly reduces the stability range of the chiral phase. Yet, the variational calculation predicts that the chiral phase  still persists for a certain restricted parameter range. With reference to the results of such variational calculation, we also perform extensive MC simulations on the lattice discretized version of the chiral GL model. In contrast to the variational results, MC gives no evidence of the chiral phase. If it exists, the stability range of the chiral phase is extremely narrow, its width being less than $0.1\%$ in the relative temperature.

The rest of the paper is organized as follows. In Sec. \ref{Model}, we describe the GL model relevant to our present study, and briefly review the previous results on the model. In Sec. \ref{Analytic}, we present the results of our variational calculation taking account of the possibility of a first-order transition. In Sec. \ref{MC_method}, we explain the details of our MC simulations. MC results are presented in Sec. \ref{Heisen} for the Heisenberg case, and in Sec. \ref{XY} for the {\it XY\/} case. Finally in Sec. \ref{Discussion}, we summarize our main results and further discuss the possibility of the chiral phase in regularly frustrated 3D spin systems. Appendices are devoted to the details of the variational calculations.


\section{\label{Model} The Chiral GL Model}

In this paper, we discus the possibility of the chiral phase in regularly frustrated 3D vector spin systems based on the following Gintzburg-Landau Hamiltonian \cite{Kawamura_1988,Kawamura_rev},
\begin{multline}
 \mathcal{H}= \frac{1}{2}\int
 d\bm{r}\Biggl\{
         \left(\bm{\nabla}\vec{a}\right)^2
         +\left(\bm{\nabla}\vec{b}\right)^2 +
         r\left(\vec{a}^2+\vec{b}^2\right) \\+u \left(\vec{a}^2 +
           \vec{b}^2\right)^2 + v \left[\left(\vec{a}\cdot \vec{b}
                                        \right)^2 -
           \vec{a}^2\vec{b}^2\right]\Biggr\},
\label{GL_eq}
\end{multline}
where $\vec{a}(\bm{r})$ and $\vec{b}(\bm{r})$ are $n$-component vector fields associated with the noncollinear spin structure at wavevectors $\pm \bm{Q}$ via
\begin{equation}
 \vec{S}(\bm{r})=\vec{a}(\bm{r})\cos(\bm{Q}\cdot\bm{r})+\vec{b}(\bm{r})\sin(\bm{Q}\cdot\bm{r}).
\label{eq-2-1}
\end{equation}
In order eq.\eqref{eq-2-1} to actually represent the noncollinear spin structure, the quartic coupling $v$ should be positive so that $\vec{a}$ and $\vec{b}$ prefer to be orthogonal to each other. Note that in order to bound the free energy, we need to limit the range of $u$ and $v$ as
\begin{equation}
 u > 0 \quad \text{and} \quad v/u < 4.
\end{equation}
This effective Hamiltonian can be derived from a microscopic spin Hamiltonian with isotropic bilinear interactions via the Hubbard-Stratonovich transformation \cite{Kawamura_1988}. In such a case, the ratio $v/u$ becomes $4/3$, while additional higher-order terms, which do not explicitly appear in (1), are also generated. Hereafter, we call the GL Hamiltonian (1) the ``chiral GL model''.

\begin{figure}
 \includegraphics[width=6cm]{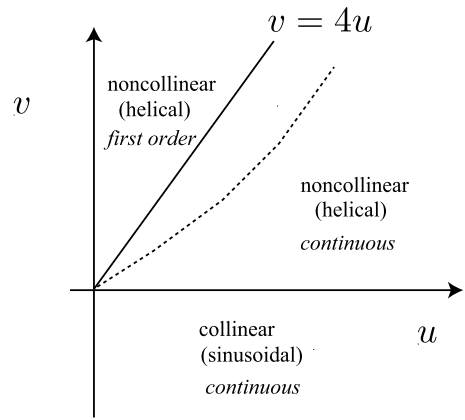}
 \caption{The mean-field phase diagram  of the chiral GL  model in the
 $(u,v)$ plane. The dotted curve is an expected boundary of a  first-order transition when fluctuations are introduced.}
 \label{fig_mean}
\end{figure}

 In the mean field approximation, a continuous transition takes place at $r=0$ \cite{Kawamura_1988}. When $v > 0$, the ordered phase is a helical magnetic state characterized by
\begin{equation}
 |\vec{a}|^2=|\vec{b}|^2= -\frac{r}{4u-v} \qquad \vec{a} \perp \vec{b}
 \qquad (0 < v < 4u).
\end{equation}
When  $v < 0$, the ordered phase is a linearly polarized  sinusoidal state characterized by
\begin{equation}
 |\vec{a}|^2+|\vec{b}|^2 = -\frac{r}{2u}\qquad \vec{a} \parallel \vec{b}
 \qquad (v < 0).
\end{equation}
When $u < 0$ or $v/u > 4$, the free energy is unstable and a higher-order term is needed to stabilize it. In such a case, the transition generally becomes of first order. The mean-field phase diagram is summarized in Fig.\ref{fig_mean}. It may be worthwhile noting that, although the mean-field approximation predicts a continuous transition for $v/u < 4$, fluctuations might change this result leading to a first-order transition, especially near the boundary $v=4u$.

 By means of a RG analysis of the chiral GL model, Kawamura found a new
 fixed point distinct  from the standard Wilson-Fisher $O(n)$ fixed
 point for certain range of the parameters \cite{Kawamura_1988,
 Kawamura_rev}. In his analysis, however, the possibility of the chiral
 phase was not considered. Since the RG expansion employed in
 ref.\cite{Kawamura_1988, Kawamura_rev} was an expansion from dimension
 four or from the many-component limit $n\to \infty$ where the chiral
 phase is never expected to occur, the chiral phase might be missed due
 to an intrinsic limitation of the method employed even if it actually
 exists in 3D in a certain parameter range. 

 Other field theoretical approaches supported the existence of a new
 fixed point  \cite{Vicari1,Vicari2}. They also performed a direct Monte Carlo
 simulation of the chiral GL model in case of $n=2$ \cite{Vicari2}, and they
 found evidence of the new universality class, although it concerned
 with a parameter range different from the target of this paper, which
 corresponded to smaller $v/u$ values.

 Recently, on the basis of a variational approximation, Onoda {\it et
 al} predicted that the chiral GL model  in 3D might exhibit a chiral
 phase characterized by $\langle \vec{a}\times \vec{b}\rangle \neq 0$
 with $\langle \vec{a}\rangle = \langle \vec{b} \rangle = \vec{0}$, if
 the quartic couplings $v$ and $u$ satisfy the relation $v/u > 4/3$
 \cite{Onoda}. Although there has not been clear evidence of the chiral
 phase in regularly frustrated 3D systems so far, the suggestion by
 Onoda {\it et al} promotes us to further examine the possible appearance of the chiral phase in the 3D chiral GL model. 

%
%

\section{\label{Analytic}Analytical consideration}

In this section, we study the ordering of the 3D chiral GL model analytically, either by the variational calculation (subsection A) or by the mapping to the nonlinear $\sigma$ model (subsection B).

\subsection{Variational Approximation}

 In this subsection,  following the analysis by Onoda {\it et al} \cite{Onoda}, we study the ordering of the 3D chiral GL model based on the variational approximation. Although some results were already reported by these authors, we will also present them for the sake of completeness. The main difference of our analysis from that of Onoda {\it et al} is that we consider the possibility of a first-order transition which was not considered by Onoda {\it et al}. In fact, a first-order transition is realized  within the variational approximation, significantly reducing the stability range of the chiral phase. 

First, we deal with the case of the Heisenberg spin ($n=3$), assuming that the macroscopic vector  chirality appears in the $z$-direction.  The variational  Hamiltonian for the chiral GL model may be given by
\begin{multline}
\mathcal{H}_0 = \frac{1}{2}\int d\bm{r}
\Bigl\{\left(\bm{\nabla}\delta\vec{a}\right)^2
+\left(\bm{\nabla}\delta\vec{b}\right)^2\\ + r_{\parallel}\left[(\delta a_x)^2+(\delta a_y)^2+(\delta
 b_x)^2+(\delta b_y)^2\right]  \\
 + r_{\perp}\left [(\delta a_z)^2+(\delta b_z)^2\right]\\
 - h_\kappa (\delta a_x\delta b_y-\delta a_y\delta b_x)\Bigr\}.
 \label{var_ham}
 \end{multline}
where $\delta \vec{a}$ and $\delta \vec{b}$ represent the deviations of the fields from their average values with respect to the variational Hamiltonian $\mathcal{H}_0$, {\it i.e.\/}, $\delta \vec{a}(\bm{r})\equiv \vec{a}(\bm{r}) -\langle \vec{a}\rangle_0$ and $\delta \vec{b}(\bm{r})\equiv \vec{b}(\bm{r}) -\langle
\vec{b}\rangle_0$, where $\langle \cdots \rangle_0$ being the average with respect to  $\mathcal{H}_0$.

 In terms of the new two-component vectors  $\vec{\alpha}$, $\vec{\beta}$, $\vec{\gamma}$ defined by
\begin{align}
 &\alpha_1 = \frac{1}{\sqrt{2}} (a_x+b_y),  &\alpha_2 &= \frac{1}{\sqrt{2}} (a_y-b_x)\notag \\
 &\beta_1 =\frac{1}{\sqrt{2}}(a_y+b_x),    &\beta_2 &= \frac{1}{\sqrt{2}}
 (-a_x+b_y)\notag\\
&\gamma_1 = a_z,&\gamma_2 &=b_z ,
\label{def_newvari}
\end{align}
the variational Hamiltonian $\mathcal{H}_0$ can be diagonalized as
\begin{multline}
 \mathcal{H}_0 = \frac{V}{2}\sum_{\bm{q}}
 \Bigl[\left(q^2+r_\parallel-\frac{h_\kappa}{2}\right)
 \delta\vec{\alpha}_{\bm{q}}\cdot
  \delta\vec{\alpha}_{\bm{-q}} \\
  +\left(q^2+r_\parallel+\frac{h_\kappa}{2}\right)
 \delta\vec{\beta}_{\bm{q}}\cdot
  \delta\vec{\beta}_{\bm{-q}} \\
  +\left(q^2+r_\perp\right)
 \delta\vec{\gamma}_{\bm{q}}\cdot
  \delta\vec{\gamma}_{\bm{-q}}
 \Bigr],
 \label{var_ham2}
\end{multline}
where $V$ represents the volume of the system.

 Let us denote $\vec{A}\equiv \langle \vec{\alpha} \rangle_0$, $\vec{B}\equiv \langle \vec{\beta} \rangle_0$, $\vec{C}\equiv \langle \vec{\gamma} \rangle_0$, with $\delta\vec{\alpha}=\vec{\alpha}-\vec{A}$, $\delta\vec{\beta}=\vec{\beta}-\vec{B}$, and $\delta\vec{\gamma}=\vec{\gamma}-\vec{C}$. Then, the spin order parameter and the vector chirality order parameter are given by
\begin{align}
 \langle \vec{a} \rangle_0&=\frac{1}{\sqrt{2}}\begin{pmatrix}
                            A_1-B_2 \\
                                              A_2+B_1\\
                                              \sqrt{2}C_1
                           \end{pmatrix},
\notag\\
 \langle \vec{b} \rangle_0&=\frac{1}{\sqrt{2}}\begin{pmatrix}
                            B_1-A_2 \\
                                              B_2+A_1\\
                                              \sqrt{2}C_2
                           \end{pmatrix},
\notag\\
 (\langle \vec{a} \times \vec{b}\rangle_0)_z &= \langle
 \vec{\alpha}^2-\vec{\beta}^2\rangle_0 \notag \\
&= \vec{A}^2-\vec{B}^2 +\frac{2}{V}\sum_{\bm{q}}\frac{h_\kappa}{(q^2+r_\parallel)^2-h_\kappa^2/4}.
\end{align}
 The chiral phase is characterized by $\vec{A}=\vec{B}=\vec{C}=\vec{0}$ and $h_\kappa \neq 0$.

 In order to determine the optimal values of the variational parameters $r_\parallel,r_\perp,h_\kappa,\vec{A},\vec{B}, \vec{C}$ within the present variational approximation, we employ the so-called Feynman inequality,
\begin{equation}
 \mathcal{F}_0 + \langle \mathcal{H}-\mathcal{H}_0\rangle_0 \ge \mathcal{F},
\label{Feynmann_ineq}
\end{equation}
where $\mathcal{F}$ ($\mathcal{F}_0$) is the free energy associated with the Hamiltonian $\mathcal{H}$ ($\mathcal{H}_0$). The optimal values of the parameters are then determined by minimizing the l.h.s of \eqref{Feynmann_ineq}. The detailed form of the l.h.s. is given in Appendix \ref {App1}.

 Let us assume $A_{1}=m \ge 0$ and other mean values are all zero in the ordered state, the vector chirality $\langle \vec{a}\times\vec{b}\rangle$ pointing along the $z$-direction. In order to avoid the ultraviolet divergence, we introduce here an upper cutoff of wavevector $\Lambda$. Various parameters are then rescaled as
\begin{align}
 \tilde{r}&= \frac{r}{\Lambda^2},&  \tilde{u}&= \frac{u}{2\pi^2\Lambda}
 ,& \tilde{v}&= \frac{v}{2\pi^2\Lambda}, \notag \\
 \tilde{r}_\parallel&= \frac{r_\parallel}{\Lambda^2},&  \tilde{r}_\perp&= \frac{r_\perp}{\Lambda^2}, &\tilde{h}_\kappa&= \frac{h_\kappa}{\Lambda^2}, \notag \\
 \tilde{m}^2&= \frac{2\pi^2m^2}{\Lambda}.
\label{scaled_var}
\end{align}
By taking the derivatives of the l.h.s. of \eqref{Feynmann_ineq} with respect to $\tilde{r}_\parallel$, $\tilde{r}_\perp$, $\tilde{h}_\kappa$ and $ \tilde{m}$, and setting them to zero, we get the following conditions for the optimal parameter values 
\begin{align}
 \tilde{r}_\parallel &= \tilde{r} +
 \left(3\tilde{u}-\frac{\tilde{v}}{4}\right)\left(\tilde{m}^2+\sigma^2_\alpha+\sigma^2_\beta\right) +2\left(\tilde{u}-\frac{\tilde{v}}{4}\right)\sigma^2_\gamma,
\label{eq_r0}\\
\tilde{r}_\perp&=r+2\left(\tilde{u}-\frac{\tilde{v}}{4}\right)\left(\tilde{m}^2+\sigma^2_\alpha
 +\sigma^2_\beta\right)
+ 4\tilde{u}\sigma^2_\gamma, \label{eq_rp}\\
 \tilde{h}_\kappa&=
 2\left(\frac{3}{4}\tilde{v}-\tilde{u}\right)\left(\tilde{m}^2+\sigma^2_\alpha-\sigma^2_\beta\right),
 \label{eq_k} \\
0&=\Biggl[
 2\left(\tilde{u}-\frac{\tilde{v}}{4}\right)\tilde{m}^2- \left(\tilde{r}_\parallel-\frac{\tilde{h}_\kappa}{2}\right)\Biggr]\tilde{m} \label{eq_m},
\end{align}
where the variances of $\alpha \sim \gamma$, $\sigma^2_\alpha \sim \sigma^2_\gamma$, are given by
\begin{align}
 \sigma^2_{\alpha}&\equiv\sum
 _{\bm{q}}\langle\delta\vec{\tilde{\alpha}}_{\bm{q}}\cdot
 \delta\vec{\tilde{\alpha}}_{-\bm{q}}\rangle_0\notag \\
 &=2\int_0^1 \frac{q^2}{q^2+\tilde{r}_\parallel-\tilde{h}_\kappa/2}dq ,\notag\\
 \sigma^2_{\beta}&\equiv\sum
 _{\bm{q}}\langle\delta\vec{\tilde{\beta}}_{\bm{q}}\cdot
 \delta\vec{\tilde{\beta}}_{-\bm{q}}\rangle_0\notag \\
 &=2\int_0^1 \frac{q^2}{q^2+\tilde{r}_\parallel+\tilde{h}_\kappa/2}dq,\notag\\
 \sigma^2_{\gamma}&\equiv\sum
 _{\bm{q}}\langle\delta\vec{\tilde{\gamma}}_{\bm{q}}\cdot
 \delta\vec{\tilde{\gamma}}_{-\bm{q}}\rangle_0\notag \\
 &=2\int_0^1 \frac{q^2}{q^2+\tilde{r}_\perp}dq.
\label{eq_sigma}
\end{align}
Note that Eq.\eqref{eq_m} has the following two types of solutions
\begin{align}
 \tilde{m}&= 0 , \\
 \tilde{m}^2& = \frac{\tilde{r}_\parallel -\tilde{h}_\kappa/2}{2(\tilde{u}-\tilde{v}/4)}.
\end{align}
The latter case corresponds to the standard helical phase, while the chiral phase corresponds to the former case.

 As discussed by Onoda {\it et al}, a solution with $\tilde{m}=0$ and $\tilde{h}_\kappa \neq 0$ is possible if $\tilde{v}$ and $\tilde{u}$ satisfy a relation
$\tilde{v}/\tilde{u} > 4/3$ \cite{Onoda}. Setting $\tilde{m}=0$ in \eqref{eq_k}, we get
\begin{equation}
 \tilde{h}_\kappa =
 \tilde{h}_\kappa\left(3\tilde{v} -4\tilde{u}\right)\int_0^1dq\frac{q^2
 }{(q^2+\tilde{r}_\parallel)^2 -\tilde{h}_\kappa^2/4}.
 \label{chiral_condition}
\end{equation}
A solution with $\tilde{h}_\kappa \neq 0$ exists only if $\tilde{v}$ is lager than $\frac{4}{3}\tilde{u}$. The transition to the chiral phase occurs when $\tilde{r}_\parallel$ is equal to $ \tilde{r}^{(c)}_\parallel $ satisfying the relation,
\begin{equation}
 \left(3\tilde{v}-4\tilde{u}\right)\int_0^1 dq
 \frac{q^2}{(q^2+\tilde{r}_\parallel^{(c)})^2}=1.
 \label{rc_p_eq}
\end{equation}
From Eq.(18), one sees that a continuous transition to the helical phase occurs at $\tilde{r}_\parallel = \tilde{h}_\kappa/2$. By substituting $\tilde{h}_\kappa = 2\tilde{r}_\parallel$ into \eqref{chiral_condition}, the value of
$\tilde{r}_\parallel=\tilde{r}_\parallel^{(s)}$ at the chiral-to-helical transition is obtained as
\begin{equation}
 \left(3\tilde{v}-4\tilde{u}\right)\int_0^1 dq \frac{1}{q^2+2\tilde{r}_\parallel^{(s)}}=1.
 \label{rs_p_eq}
\end{equation}
From Eqs.\eqref{rc_p_eq} and \eqref{rs_p_eq}, one can show  that $\tilde{r}_\parallel^{(s)}$ is always greater than $\tilde{r}_\parallel^{(c)}$. The region $\tilde{r}_\parallel^{(c)} \le \tilde{r}_\parallel \le  \tilde{r}_\parallel^{(s)}$ then corresponds to  the chiral phase.

In order to discuss how the ordering proceeds when the ``temperature'' $\tilde{r}$ is varied, we need to calculate the ``transition temperatures'' $\tilde{r}_c$ and $\tilde{r}_s$ corresponding to $\tilde{r}_\parallel^{(c)}$ and $\tilde{r}_\parallel^{(s)}$.  Although the inequality $\tilde{r}_\parallel^{(c)} \le \tilde{r}_\parallel^{(s)}$ always holds, the relation between the corresponding $\tilde{r}_c$ and $\tilde{r}_s$ is not trivial depending on the values  $\tilde{u}$ and $\tilde{v}$. For a fixed ratio $\tilde{v}/\tilde{u}$, we can show that there is a critical value $\tilde{u}_c$ such that $\tilde{r}_s > \tilde{r}_c$ for $\tilde u < \tilde u_c$ and $\tilde{r}_s < \tilde{r}_c$  for $\tilde u > \tilde u_c$ (details are given in Appendix \ref{App1}). Note that only  the latter situation means the existence of the chiral phase. Indeed, in the case of $\tilde{u} < \tilde{u}_c$, a first-order transition from the paramagnetic phase to helical phase occurs directly without an intermediate chiral phase.

\begin{figure}
 \includegraphics[width=5cm]{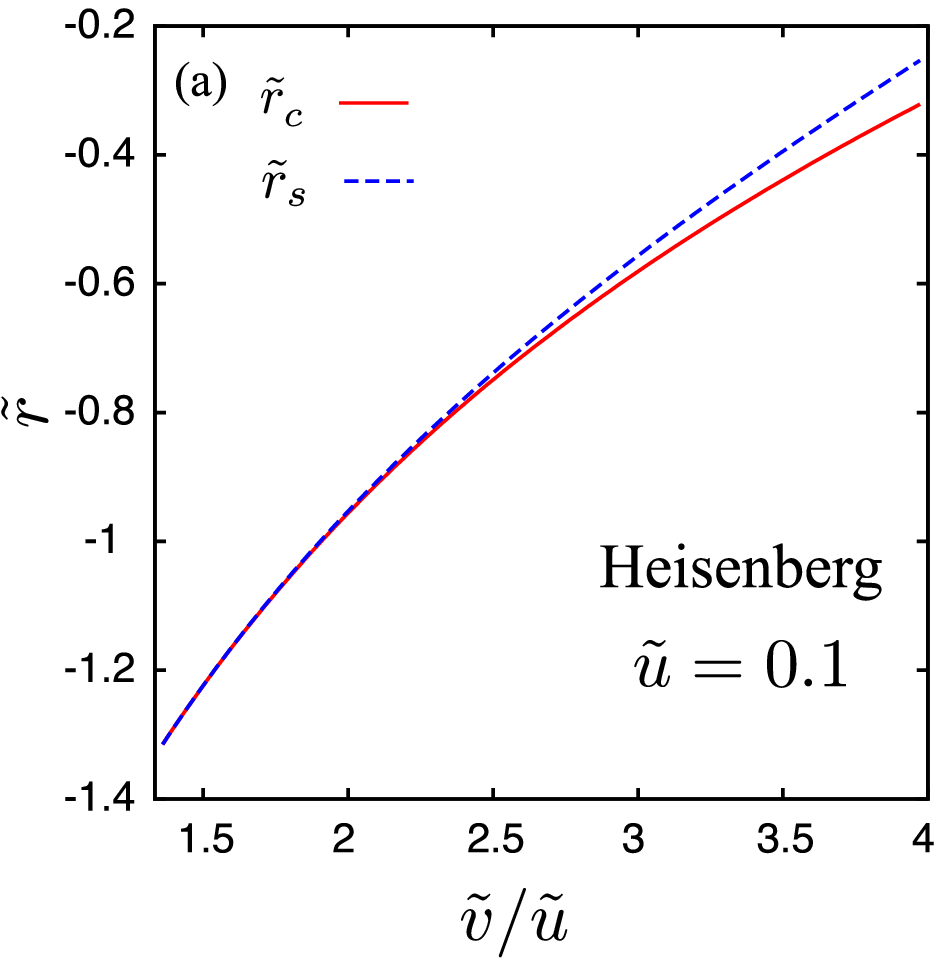}
 \includegraphics[width=5cm]{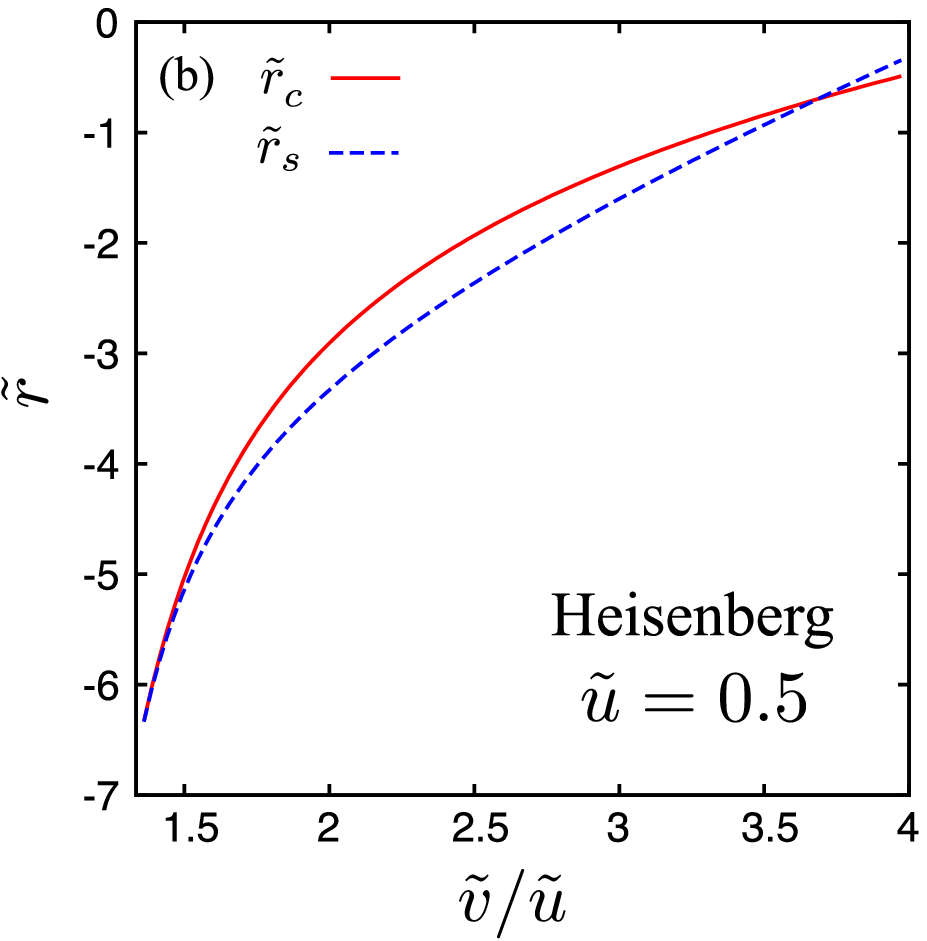}
 \caption{(Color online) The ``transition temperatures'' $\tilde{r}_c$ and $\tilde{r}_s$ as functions of $\tilde{v}/\tilde{u}$ for the case of (a) $\tilde{u}=0.1$, and (b) $\tilde{u}=0.5$, where $\tilde{r}_c$ and $\tilde{r}_s$ are the para-to-chiral and the chiral-to-helical continuous transition points.}
\label{fig_an1}
\end{figure}

 Fig. \ref{fig_an1} exhibits $\tilde{r}_c$ and $\tilde{r}_s$ as a function of $\tilde{v}/\tilde{u}$ for the case of $\tilde{u}=0.1$ and $0.5$. For $\tilde{u}=0.1$, an inequality $\tilde{r}_c < \tilde{r}_s$ is satisfied for $ 4/3 \lesssim\tilde{v}/\tilde{u} < 4$ (there may be a tiny region of $\tilde{r}_c > \tilde{r}_s$ in a close vicinity of $\tilde{v}/\tilde{u} = 4/3$ ). The chiral phase does not exist for this small value of $\tilde{u}$. By contrast, for $\tilde{u}=0.5$, an inequality $\tilde{r}_c >\tilde{r}_s$ is satisfied  for $ 4/3 \le \tilde{v}/\tilde{u} \lesssim 3.68$, and the chiral phase exists for this large $\tilde{u}$.

 So far, we have considered only the case of $\tilde{m}=0$. However, if a first-order transition really occurs, a continuous transition to the chiral phase at $\tilde{r}_c$ might be interrupted by such a first-order transition, and we need to examine the case of $\tilde{m}\neq 0$ simultaneously, choosing the state giving the lower free energy within the variational approximation. This point has not been examined in Ref.\cite{Onoda}. The first-order transition point is located by comparing the free energies (the l.h.s. of \eqref{Feynmann_ineq}) of the  $\tilde{m}=0$ and $\tilde{m}\neq 0$ solutions. The explicit form of the free energy is given in appendix \ref{App1}.

 In Fig.\ref{fig_an2}, we show the numerically calculated transition temperatures as functions of $\tilde{u}$ for several values of $\tilde{v}/\tilde{u}$. The first-order transition temperature, $\tilde{r}_1$, is usually greater than $\tilde{r}_s$. It means that the transition to the helical phase is of first-order within this variational approximation. Recall here that, even in the typical $\phi^4$
 model describing a ferromagnet, the same variational approximation predicts an artificial first-order transition contrary to the reality. Therefore, the first-order nature of the transition might also be an artifact of the variational  approximation employed here.
\begin{figure}
 \includegraphics[width=5cm]{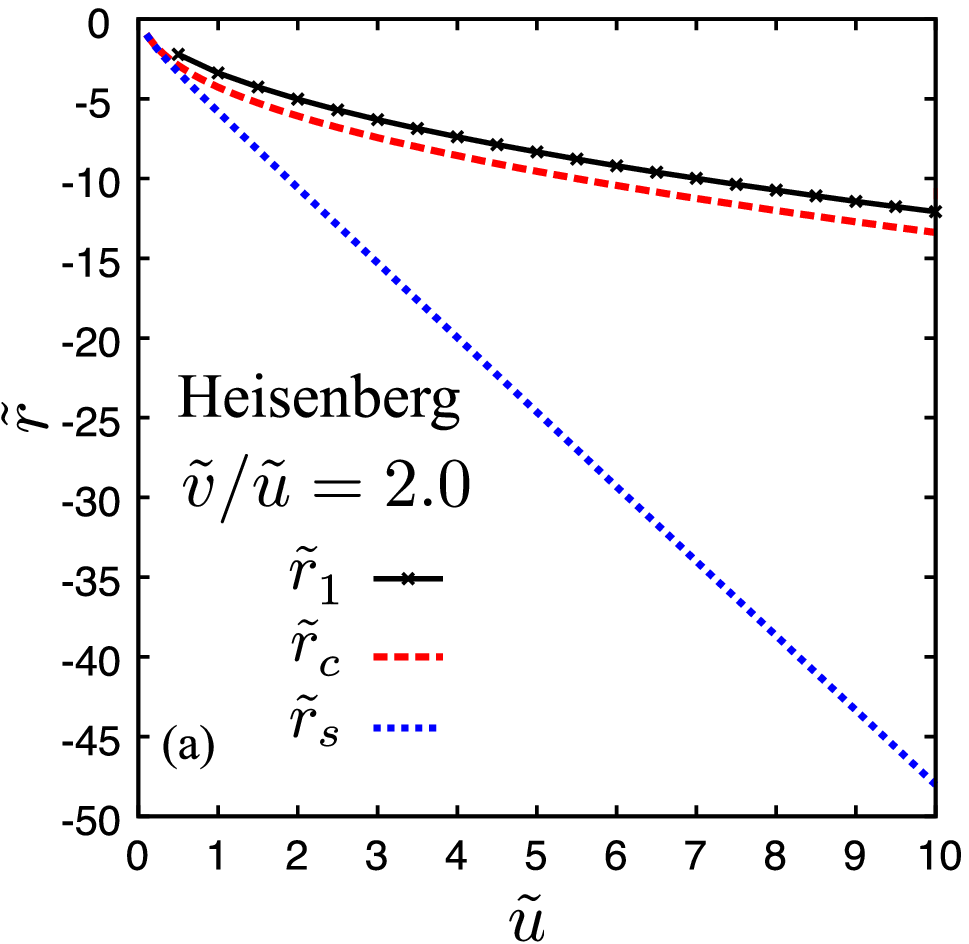}
 \includegraphics[width=5cm]{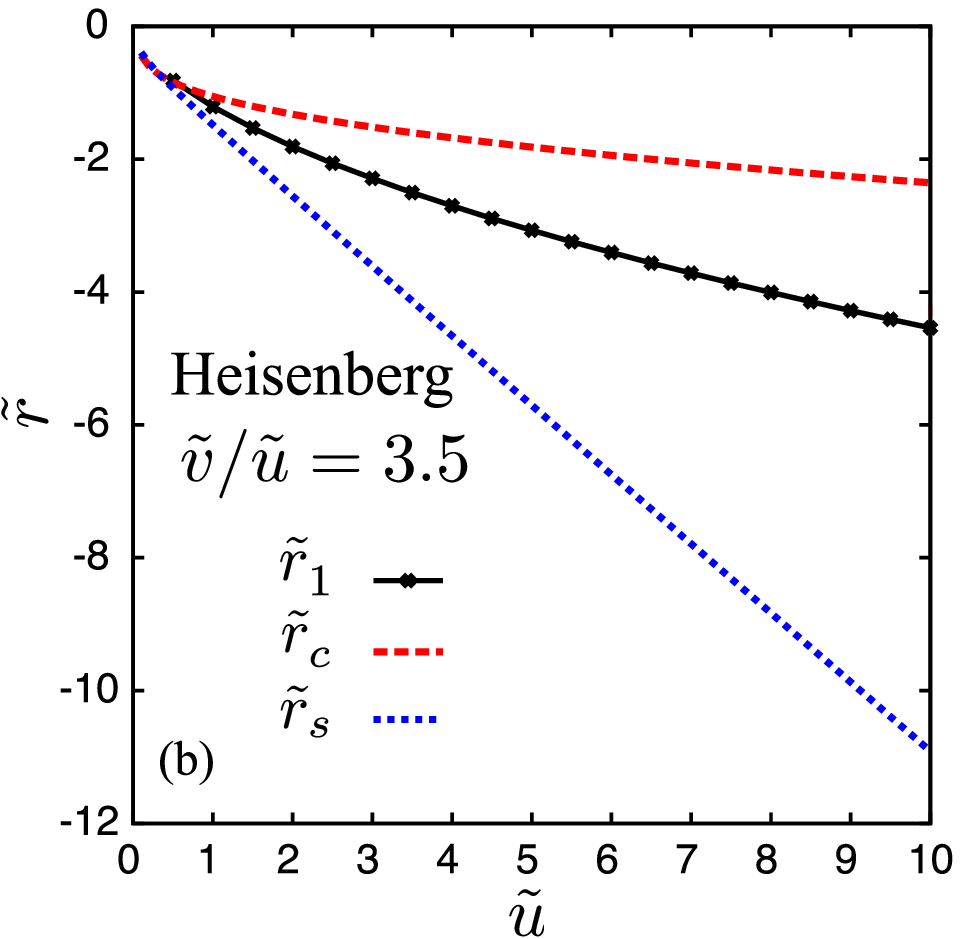}
 \caption{(Color online) The ``transition temperatures'' $\tilde{r}_c$,$\tilde{r}_s$
 and $\tilde{r}_1$ as functions of  $\tilde{u}$ for the case of (a)
 $\tilde{v}/\tilde{u}=2.0$, and (b) $\tilde{v}/\tilde{u}=3.5$, where
 $\tilde{r}_c$ and $\tilde{r}_s$ are the para-to-chiral and the
 chiral-to-helical continuous transition points which would occur if the
 possibility of a first-order transition would be neglected, and
 $\tilde{r}_1$ is the para-to-helical (or chiral-to-helical) first-order transition point.}
\label{fig_an2}
\end{figure}
In case of $\tilde{v}/\tilde{u}=2.0$, we see from Fig.3(a) that $\tilde{r}_1> \tilde{r}_c >\tilde{r}_s$, at least for $\tilde{u}<10$. It means that a continuous transition to the chiral phase is not realized, and alternatively, a first-order transition to the helical phase occurs at a higher temperature. On the other hand, an inequality $\tilde{r}_1 < \tilde{r}_c$ is satisfied for $\tilde{v}/\tilde{u}=3.5$. In this case, with decreasing the temperature a continuous transition to the chiral phase occurs first, and then, the system goes into the helical phase through a first-order transition at $\tilde{r}=\tilde{r}_1$.

\begin{figure*}
 \includegraphics[width=7cm]{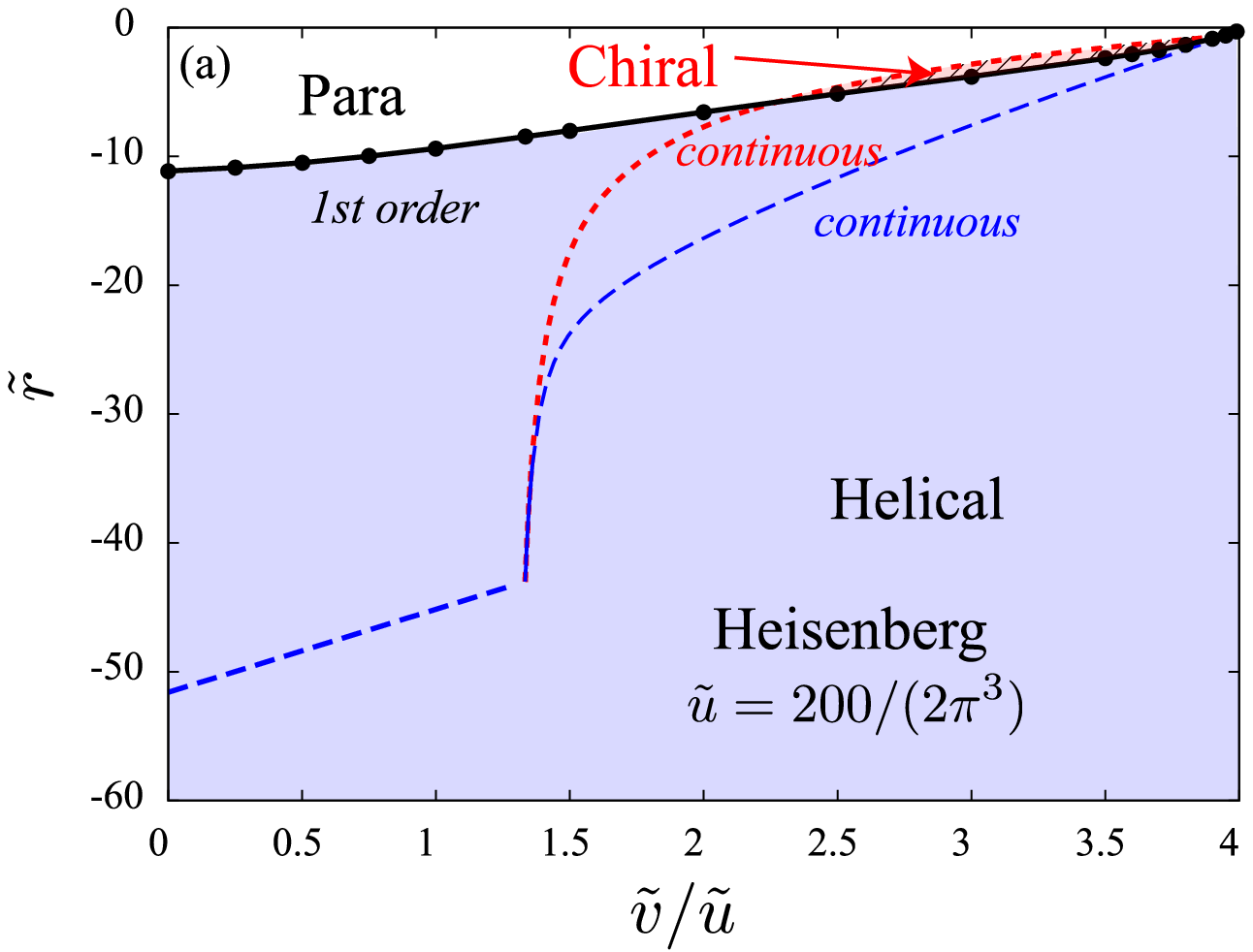}
 \includegraphics[width=7cm]{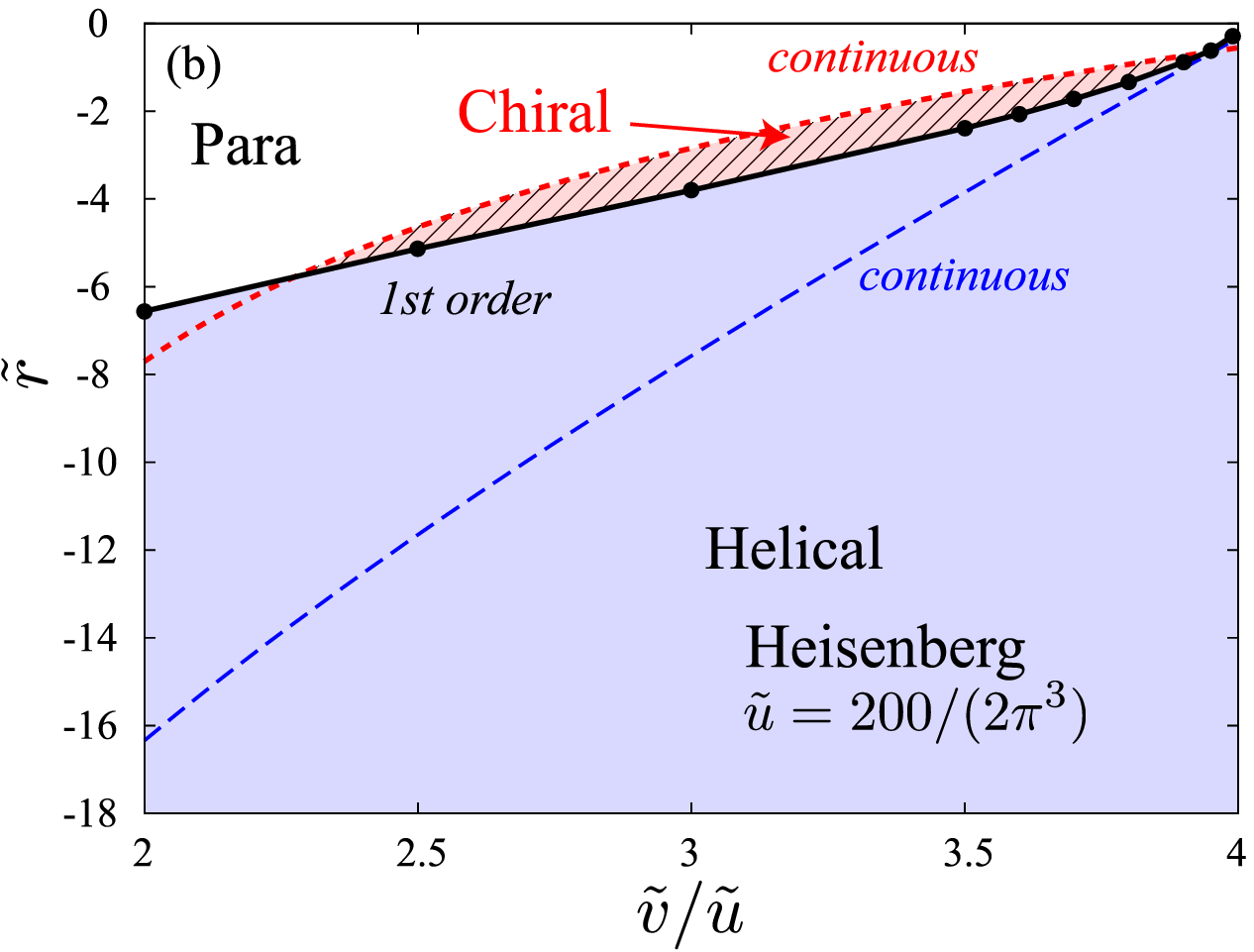}
 \caption{(Color online) Phase diagram of the Heisenberg ($n=3$) chiral
 GL model in the  $(\tilde{r},\tilde{v}/\tilde{u})$ plane for
 $\tilde{u}=200/(2\pi^3)$. The right figure is an enlarged view of the
 vicinity of the chiral phase. The red hatched (blue filled) area represents the chiral (helical) phase. The dotted and dashed curves represent continuous transition lines which would occur if the possibility of a first-order transition would be neglected in the analysis. The chiral phase predicted by Onoda {\it et al} occupies the region between these two curves\cite{Onoda}. The solid curve represents the first-order transition line from the paramagnetic phase to the helical phase which is newly found in this work. The stability range of the chiral phase is largely reduced compared with that reported by Onoda {\it et al}.
}
\label{fig_phase}
\end{figure*}

 The phase diagram of the 3D chiral GL model in the $(\tilde{r},\tilde{v}/\tilde{u})$ plane is shown in Fig.\ref{fig_phase} for lager $\tilde{u}$. The dotted and the dashed curves represent continuous transitions which would occur if we would ignore the possibility of a first-order transition. The chiral phase predicted by Onoda {\it et al} occupies the region between these two curves\cite{Onoda}. The solid curve represents the first-order transition from the paramagnetic phase to the helical phase which is newly found in this work. Although the range of the chiral phase is largely reduced due to the first-order transition, the chiral phase still persists for $\tilde{v}/\tilde{u} \gtrsim 2.5$ in the case of $\tilde{u}=200/(2\pi^3) \sim 3.22$.

 Thus, within the variational approximation, the chiral phase appears for sufficiently large $\tilde{u}$ and $\tilde{v}/\tilde{u}$. The range of the chiral phase becomes wider for larger $\tilde{u}$, as can be seen from Fig.\ref{fig_an2}. 

 For the case of the {\it XY\/} spin ($n=2$), the same analysis as was done in the Heisenberg case can also be performed by simply removing the $r_\perp$ term from the variational Hamiltonian \eqref{var_ham} and neglecting the $\vec{\gamma}$ term: See appendix \ref{App1}. The results are qualitatively the same as those of the Heisenberg case, only the values of the transition temperatures being different. Hence, we conclude that the chiral phase exists within the variational approximation for sufficiently large $\tilde{u}$ and $\tilde{v}/\tilde{u}$ even for the {\it XY\/} case.

\subsection{The $u \to \infty$ limit}

 In this subsection, we consider the behavior of the chiral GL model in the limit of $u \to \infty$ with keeping $r/u$ and $v/u$ constant. In this limit, the model described by the Hamiltonian \eqref{GL_eq} reduces to the NL$\sigma$ model \cite{NLS} given by, 
\begin{equation}
 \mathcal{H}_{\mathrm{NL\sigma}}=\frac{1}{T}\int d\bm{r}\left\{(\bm{\nabla}\vec{a})^2
         +(\bm{\nabla}\vec{b})^2
         +R[\bm{\nabla(\vec{a}\times\vec{b}})]^2\right\}, 
\label{sigma_eq}
\end{equation}
with conditions
\begin{equation}
 |\vec{a}|^2=|\vec{b}|^2=1 , \qquad \vec{a}\perp\vec{b}.
\end{equation}

From a simple saddle point analysis, one can derive the relation between the variables $T$ and $R$ of the NL$\sigma$ model and the parameters of the chiral GL model. Indeed, the chiral GL model Hamiltonian \eqref{GL_eq} can be rewritten
by introducing $r^\ast=r/u$ and $v^\ast=v/u$ as
\begin{multline}
 \mathcal{H}_c= \frac{1}{2}\int
 d\bm{r}\left\{(\bm{\nabla}\vec{a})^2
         +(\bm{\nabla}\vec{b})^2\right\}\\
 +\frac{u}{2}\int d\bm{r} \Biggl\{
         r^\ast\left(\vec{a}^2+\vec{b}^2\right) +
         \left(\vec{a}^2 +
           \vec{b}^2\right)^2 \\
 + v^\ast \left[\left(\vec{a}\cdot \vec{b}
                                        \right)^2 -
           \vec{a}^2\vec{b}^2\right]\Biggr\} .
\label{GL_eq2}
\end{multline}
Note that the second term of the r.h.s. becomes much larger compared with the first term in the limit of $u\to \infty$. Therefore, one can apply the saddle point approximation to the partition function associated with \eqref{GL_eq2},
\begin{equation}
 \mathcal{Z}_c =
 \int\mathcal{D}\vec{a}(\bm{r})\int\mathcal{D}\vec{b}(\bm{r})
 \exp(-\mathcal{H}_c).
\end{equation}
Since the second term of \eqref{GL_eq2} contains only on-site interactions, one  can easily evaluate the minimization conditions for the second term as
\begin{align}
 |\vec{a}|^2=|\vec{b}|^2&= -\frac{r^\ast}{4-v^\ast} \\
 \vec{a} &\perp\vec{b}.
\end{align}

 In the limit of $u \to \infty$, the functional integral of the partition  function is approximated by an integral within a subspace constrained  by the above conditions, yielding
\begin{align}
 \mathcal{Z}_c &\sim
 \iint\limits_{|\vec{a}|^2=|\vec{b}|^2=
 1,\vec{a}
 \perp\vec{b}}\mathcal{D}\vec{a}(\bm{r})\mathcal{D}\vec{b}(\bm{r})
 \notag \\
& \qquad \qquad \times  \exp\left\{-\frac{-r^\ast}{8-2v^\ast}\int
 d\bm{r}\left[(\bm{\nabla}\vec{a})^2
         +(\bm{\nabla}\vec{b})^2\right]\right\},
\end{align}
where we rescale the fields $\vec{a}(\bm{r})$ and $\vec{b}(\bm{r})$ so that they satisfy $|\vec{a}|^2=|\vec{b}|^2=1$. Thus, one sees that the chiral GL model in the $u\rightarrow \infty$ limit reduces to the NL$\sigma$ model with the correspondence
\begin{align}
 \label{NLS_relation}
 \frac{1}{T} &= \frac{-r}{8u-2v} , \\
 R&=0.
\end{align}
It also suggests that, in the limit of $u \to \infty$, the properties of the chiral GL model is independent of the value of $v/u$, depending only on the scaled parameter $T$ given by Eq.\eqref{NLS_relation}.

Note that the chiral GL model corresponds to the $R=0$ sector of the NL$\sigma$ model. Although David {\it et al} predicted that the stable chiral phase existed for sufficiently large $R$ \cite{NLS}, there has been no report of the chiral phase for $R=0$ \cite{NLS2,NLS3,NLS4,NLS_XY}. It indicates  that for sufficiently large $u$ there is very little chance for the chiral phase to be stabilized.

 By combining this result with that obtained from our variational analysis in the previous subsection, it seems most natural to expect that the chiral phase has the highest chance to be stabilized for intermediate values of $u$ and for larger values of $v/u$. Hence, in the following section, we perform extensive MC simulations focusing on such a parameter region in search for the possible chiral phase.

\section{\label{MC_method} Monte Carlo Simulation}
\subsection{Method}

 In this and following sections, we investigate the ordering properties of the chiral GL model numerically by MC simulations. For this purpose, the model is discretized on a 3D simple cubic lattice with lattice constant $\epsilon$ as
\begin{multline}
 \mathcal{H}_\epsilon = \frac{1}{2}\sum_i\epsilon^3 \Biggl\{  \sum_{\mu}\left(\frac{\vec{a}_{i+\mu} -\vec{a}_i}{\epsilon}\right)^2+  \sum_{\mu}\left(\frac{\vec{b}_{i+\mu} -\vec{b}_i}{\epsilon}\right)^2
 \\
 + r\left(\vec{a}_i^2 + \vec{b}_i^2\right) + u\left(\vec{a}_i^2 +
 \vec{b}_i^2\right)^2
 +v\left[\left(\vec{a}_i\cdot\vec{b}_i\right)^2
                       -\vec{a}_i^2\vec{b}_i^2\right]\Biggr\},
\label{Hamiltonian_epsilon}
\end{multline}
where $\vec{a}_i$ and $\vec{b}_i$ are $n$-component vectors at the site $i$, while $i+\mu$ represents the nearest-neighbor site of $i$ in the $\mu$-direction
($\mu=x,y,z$).  We introduce rescaled parameters as
\begin{align}
 r^\prime &= \epsilon^2r \label{r_scale}\\
 u^\prime &= \epsilon u \label{u_scale}\\
 v^\prime &= \epsilon v \label{v_scale}\\
 \vec{a}_i^\prime &= \epsilon^{\frac{1}{2}}\vec{a}_i\\
 \vec{b}_i^\prime &= \epsilon^{\frac{1}{2}}\vec{b}_i.
\end{align}
The $\epsilon$ dependence of the Hamiltonian \eqref{Hamiltonian_epsilon} is then removed as
\begin{multline}
 \mathcal{H} = \frac{1}{2}\sum_i \Biggl\{
 \sum_{\mu}\left(\vec{a}_{i+\mu}^\prime -\vec{a}^\prime_i\right)^2+
 \sum_{\mu}\left(\vec{b}_{i+\mu}^\prime -\vec{b}^\prime_i\right)^2
 \\
 + r^\prime\left[(\vec{a}^\prime_i)^2 +
 (\vec{b}^\prime_i)^2 \right]+
 u^\prime\left[(\vec{a}^\prime_i)^2 +
 (\vec{b}^\prime_i)^2\right]^2 \\
 +v^\prime\left[\left(\vec{a}^\prime_i\cdot\vec{b}^\prime_i\right)^2
                       -(\vec{a}^\prime_i)^2
 (\vec{b}^\prime_i)^2\right]\Biggr\}.
\label{lattice_eq}
\end{multline}

 Such discretization and scaling procedure just corresponds to the cutoff procedure made for the continuum model (1). Setting $\Lambda = \pi/\epsilon$, the parameters $(r^\prime, u^\prime, v^\prime)$ in eqs.\eqref{r_scale}-\eqref{v_scale} can be related to  $(\tilde r, \tilde u, \tilde v)$ in eq.\eqref{scaled_var} as
\begin{align}
 \tilde{r}&\leftrightarrow \frac{r^\prime}{\pi^2}, & \tilde{u}&\leftrightarrow \frac{u^\prime}{2\pi^3}, & \tilde{v}&\leftrightarrow \frac{v^\prime}{2\pi^3}.
\label{param_relation}
\end{align}
Note that a proportionality coefficient $1/(2\pi^3)\simeq 0.016$ means that $\tilde{u}=1$ corresponds to a rather large value of $u^\prime \simeq 62$.

We perform extensive MC simulations on the lattice chiral GL model
described by \eqref{lattice_eq} for both cases of the Heisenberg spin
($n=3$) and the {\it XY\/} spin ($n=2$) by using the standard Metropolis
method. We consider $r^\prime$ as the ``temperature'' and the
simulation is performed with the statistical weight of
$\exp(-\mathcal{H})$ at each ``temperature''. The lattice is a $L\times L \times L$ simple cubic lattice with
$8 \le L \le 60$. Periodic boundary conditions are imposed in all
directions. In updating $\vec{a}^\prime_i$ and $\vec{b}^\prime_i$
vectors, we adopt 
the polar coordinate in spin space and apply the type of Metropolis
updating where appropriate windows are set for the proposed values of
the variables so that the acceptance ratio becomes  $0.25 \sim
0.6$. When we simulate extreme cases where $u^\prime$, $v^\prime$ and
$r^\prime$ are much larger than unity, these adjustment procedures based
on appropriate polar coordinate turn out to be efficient for 
thermalization. In case of the {\it XY\/} spin ($n=2$), we also try the
exchange of the directions of the $\vec{a}^\prime_i$ and
$\vec{b}^\prime_i$ vectors according to  the Metropolis rule, which
turns out to be efficient in relaxing the chirality vector $\vec{a}_i
\times \vec{b}_i$. 

Typically, our single MC run contains $6\times 10^6$ MC steps per spin
(MCS) at each temperature $r^\prime$. In most cases, a gradual cooling
protocol is employed. For large systems and near the transition temperature,
we perform longer runs of up to $\sim 1.8\times 10^7$ MCS. In
calculating physical quantities, initial ($1 \sim 3) \times 10^6$ MCS are
discarded for thermalization, and averages are calculated over
subsequent  ($3 \sim 5) \times 10^6$ MCS.  Error bars are estimated by making $3\sim 5$ independent runs at each temperature and lattice size.

\subsection{Physical quantities}

In this subsection, we introduce various physical quantities measured in our MC simulation.

We define the ``specific heat'' as the variance of the energy per spin,
\begin{equation}
 C \equiv \frac{1}{N} \left( \langle \mathcal{H}^2\rangle - \langle \mathcal{H}\rangle^2 \right ).
\label{def_specificheat}
\end{equation}
Note that in the present model the specific heat defined by \eqref{def_specificheat} is not equivalent to the one defined by the ``temperature'' derivative of the energy, $\frac{d E}{d r^\prime}$, where $E=\frac{1}{N}<\mathcal{H}>$ is the internal energy per spin,  although both quantities are expected to exhibit  similar singular behaviors at the transition.

In order to measure the spin order, we define the spin order parameter by
\begin{equation}
 M \equiv \frac{1}{N}\sqrt{\langle (\sum_i\vec{a}^\prime_i)^2 + (\sum_i
                        \vec{b}^\prime_i )^2 \rangle}.
\end{equation}
We also define the chiral order parameter by
\begin{equation}
 \kappa \equiv  \frac{1}{N}\sqrt{\langle (\sum_i \vec{a}^\prime_i \times
                              \vec{b}^\prime_i)^2 \rangle}.
\end{equation}
The finite-size correlation lengths of the spin and of the chirality are defined on the basis of the Ornstein-Zernike form of the correlation function by
\begin{equation}
 \xi_{\alpha} \equiv
 \frac{1}{|\bm{q_m}|}\sqrt{\frac{C_\alpha(\bm{0})}{C_\alpha(\bm{q}_m)}-1},
\end{equation}
where $\alpha$ stands for either the spin ($s$) or the chirality ($c$), $C_\alpha$ being the Fourier transform of the spatial correlation function,
\begin{align}
 C_s(\bm{q})&\equiv \langle |\vec{a}^\prime_{\bm{q}}|^2
 +|\vec{b}^\prime_{\bm{q}}|^2 \rangle, \\
 C_c(\bm{q})&\equiv \langle |(\vec{a}^\prime\times \vec{b}^\prime)_{\bm{q}} |^2
 \rangle,
\end{align}
where $\vec{a}^\prime_{\bm{q}}$, $\vec{b}^\prime_{\bm{q}}$ and
$(\vec{a}^\prime\times \vec{b}^\prime)_{\bm{q}}$ represent the Fourier
transform of $\vec{a}^\prime_i$, $\vec{b}^\prime_i$, and
$\vec{a}^\prime_i\times\vec{b}^\prime_i$, respectively. In our
simulation, we take $\bm{q}_m=(2\pi/L,0,0)$ corresponding to one of the minimum
wavevectors compatible with periodic boundary conditions.

 We also measure the Binder ratios of the spin and of the chirality. For the spin, we have
\begin{equation}
 g_s \equiv \left[D_1^{(s)} - D_2^{(s)}\left(\frac{\langle
            (\sum_i\vec{a}^\prime_i)^4\rangle}{\langle (\sum_i
            \vec{a}^\prime_i)^2\rangle^2} +\frac{\langle
            (\sum_i\vec{b}^\prime_i)^4\rangle}{\langle (\sum_i
            \vec{b}^\prime_i)^2\rangle^2}\right)\right ],
\end{equation}
while, for the chirality, we have
\begin{equation}
 g_c \equiv \left[D_1^{(c)} - D_2^{(c)}\frac{\langle
            (\sum_i\vec{a}^\prime_i\times\vec{b}^\prime_i)^4\rangle}{\langle (\sum_i
            \vec{a}^\prime_i\times\vec{b}^\prime_i)^2\rangle^2}\right ].
\end{equation}
The coefficients $D_1^{(\alpha)}$ and $D_2^{(\alpha)}$ ($\alpha = s ,c$) are determined so that, in the thermodynamic limit, $g_\alpha$ vanishes in the high-temperature phase and gives unity in the ordered phase. In the Heisenberg case ($n=3$), one has $D_1^{(s)}=5/2$, $D_2^{(s)}=3/4$, $D_1^{(c)}=5/2$  and $D_2^{(c)}=3/2$, while for the {\it XY\/} case ($n=2$), one has $D_1^{(s)}=2$, $D_2^{(s)}=1$, $D_1^{(c)}=3/2$  and $D_2^{(c)}=1/2$.

\section{Simulation Results: Heisenberg ($n=3$) case \label{Heisen}}

 In this section, we present the results of our MC simulation for the Heisenberg case ($n=3$). As a typical example, we deal with the case of $u^\prime=200$ and $v^\prime/u^\prime=3.5$ here. For these parameters, the variational calculation of section \ref{Analytic} predicts that the chiral phase is stabilized in a relatively wide temperature range between $r^\prime_s\simeq-23.6$ and $r^\prime_c\simeq-15.5$. The relative difference between $r_c^\prime$ and $r_s^\prime$, a measure of the width of the chiral phase, might be given by
\begin{equation}
\delta r^\prime \equiv 2\frac{r_c^\prime-r_s^\prime}{|r_c^\prime+r_s^\prime|}.
\end{equation}
The variational calculation predicts a rather large value of $\delta r^\prime \simeq 0.42$ for these parameter values.

\begin{figure}
 \includegraphics[width=7cm]{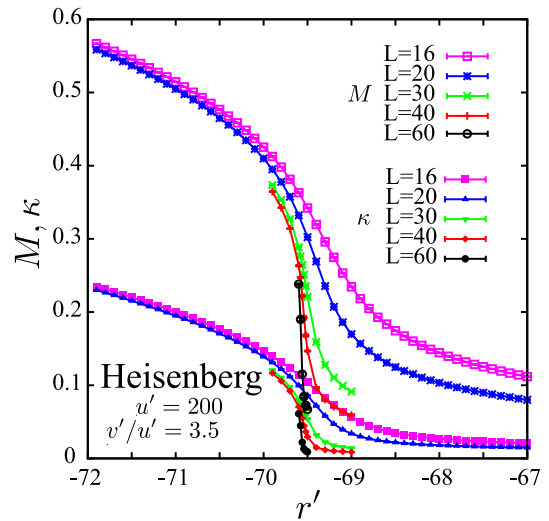}
\caption{(Color online) The temperature ($r^\prime$) dependence of the spin and the chiral order parameters of the Heisenberg ($n=3$) chiral GL model for various sizes. The Hamiltonian parameters are set $u^\prime=200$ and $v^\prime/u^\prime=3.5$.}
\label{fig1}
\end{figure}

 First, we show in Fig.\ref{fig1} the temperature ($r^\prime$) dependence of the spin and the chiral order parameters. On decreasing the temperature, both order parameters rise up sharply around $r^\prime \simeq -69.6$, implying the occurrence of a phase transition around this temperature as can be seen from the figure. Each order parameter rises up at mutually close temperatures such that their relative difference is much smaller than the one predicted from the variational calculation.

\begin{figure}
 \includegraphics[width=7cm]{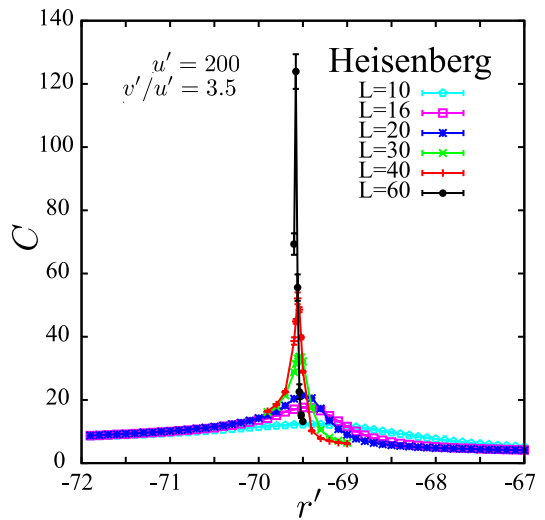}
\caption{(Color online) The temperature ($r^\prime$) dependence of the specific heat of the Heisenberg ($n=3$) chiral GL model. The Hamiltonian parameters are set $u^\prime=200$ and $v^\prime/u^\prime=3.5$.}
\label{fig2}
\end{figure}

\begin{figure}
 \includegraphics[width=7cm]{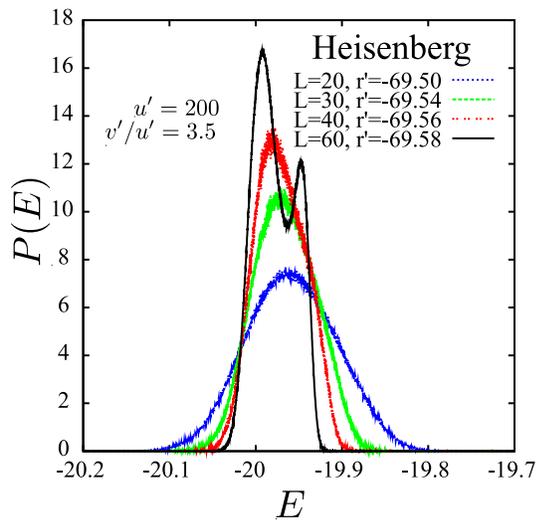}
\caption{(Color online) The energy distribution at the specific-heat peak temperature  of the Heisenberg ($n=3$) chiral GL model for various sizes. The Hamiltonian parameters are set $u^\prime=200$ and $v^\prime/u^\prime=3.5$.}
\label{fig_PE}
\end{figure}

 Fig.\ref{fig2} exhibits the temperature ($r^\prime$) dependence of the specific heat. It exhibits only a single peak with no evidence of successive transitions. Note that the peak height of our largest size $L=60$ is much larger than that of other sizes. This may be a signature of a weak first-order transition occurring in the thermodynamic limit. In fact, as shown in fig.\ref{fig_PE}, the energy distributions for $L=60$ shows double peaks characteristic of a first-order transition at the peak temperature, although the energy distributions of smaller sizes show only a single peak. In view of the fact that the parameter value studied here $v^\prime/u^\prime=3.5$ is close to the mean-field tricritical line $v^\prime/u^\prime=4$, the occurrence of a weak first-order transition seems consistent with the previous RG observation, since fluctuations extend the region of first-order transition.

 From the data of the order parameters and of the specific heat, we now expect that, even if the chiral phase exists, it is limited to a very narrow temperature range, never spreading as wide as  $\delta r\simeq 42\%$ predicted from the variational theory.  

\begin{figure}
 \includegraphics[width=7cm]{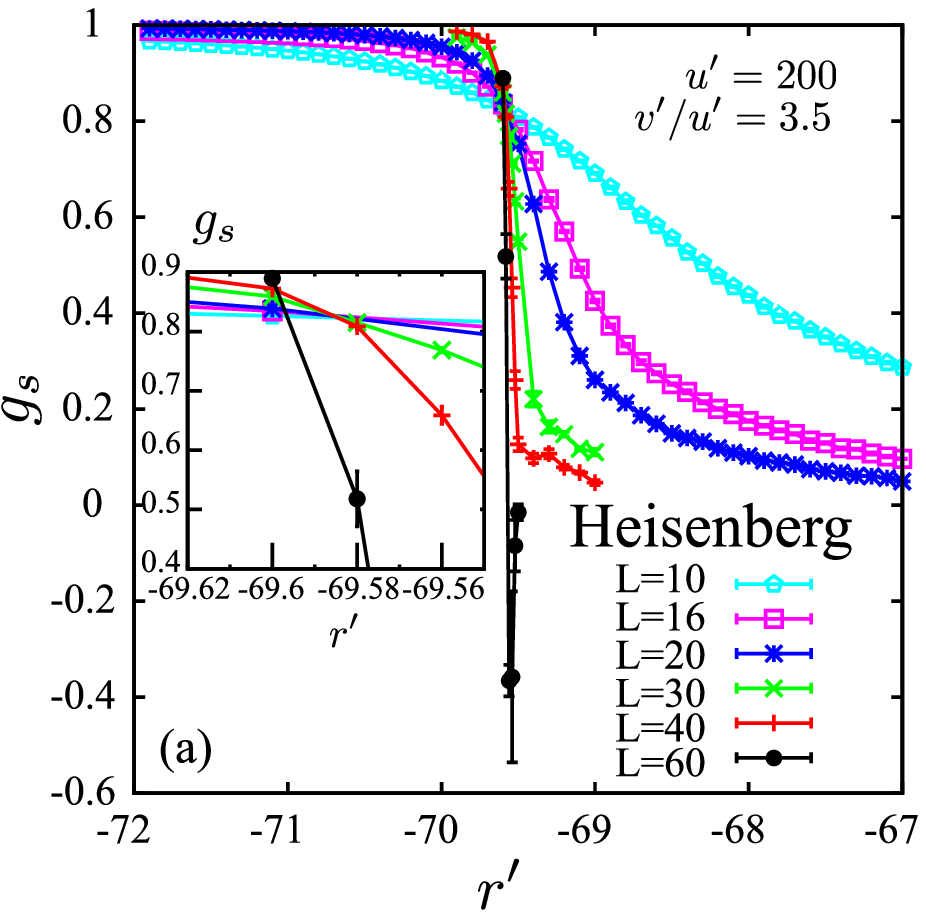}
 \includegraphics[width=7cm]{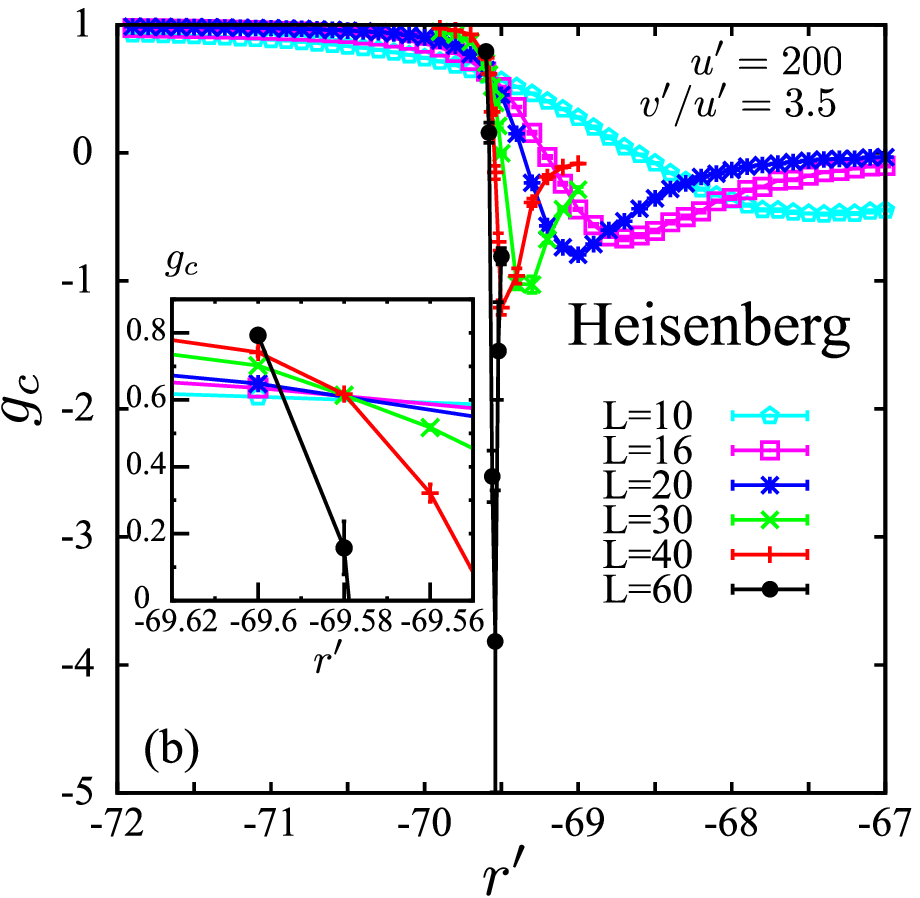}
\caption{(Color online) The temperature ($r^\prime$) dependence of the Binder ratio for the spin (a), and for the chirality (b),  of the Heisenberg ($n=3$) chiral GL model. The Hamiltonian parameters are set $u^\prime=200$ and $v^\prime/u^\prime=3.5$. Insets are enlarged views of the transition region.}
\label{fig3}
\end{figure}

\begin{figure}
 \includegraphics[width=7cm]{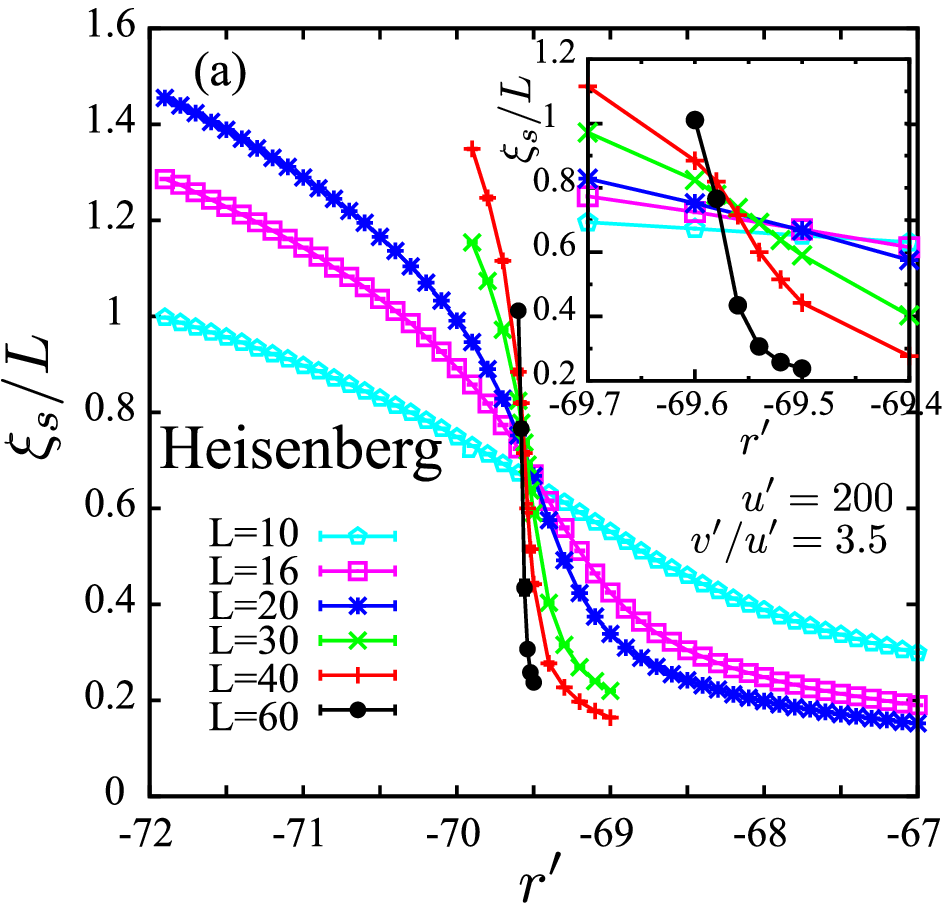}
 \includegraphics[width=7cm]{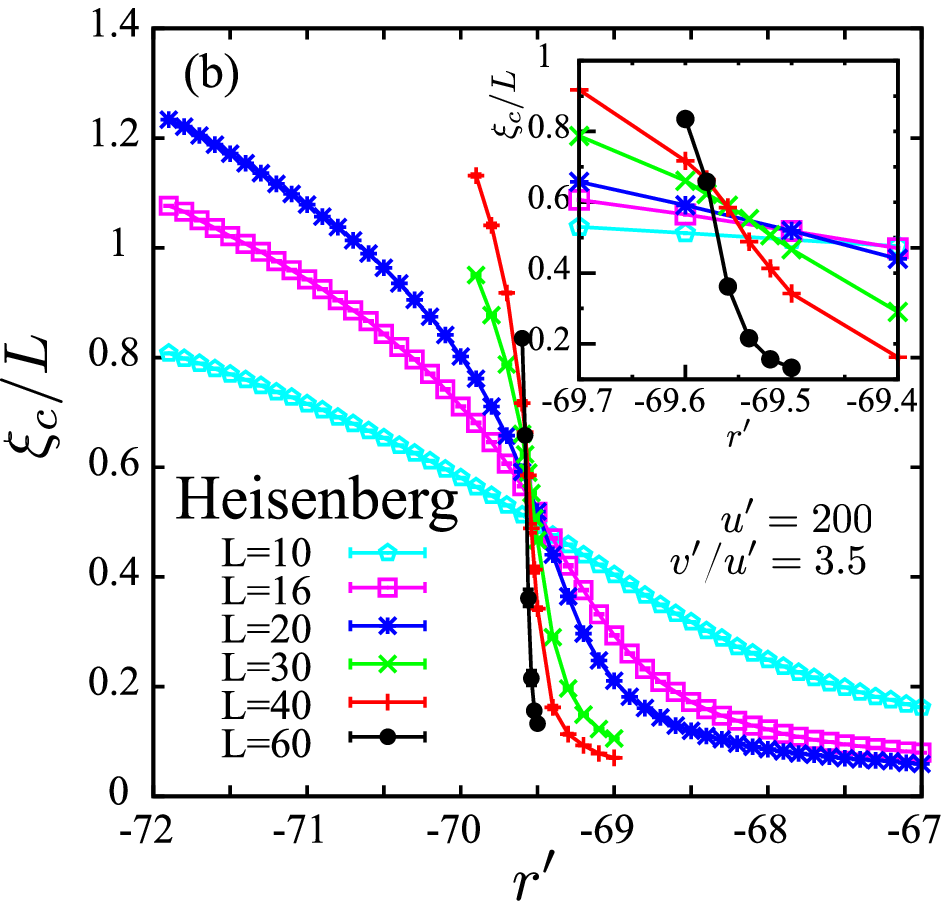}
\caption{(Color online) The temperature ($r^\prime$) dependence of the correlation-length ratio for the spin (a), and for the chirality (b), of the Heisenberg ($n=3$) chiral GL model. The Hamiltonian parameters are set $u^\prime=200$ and $v^\prime/u^\prime=3.5$. Insets are enlarged views of the transition region.}
\label{fig4}
\end{figure}

 In order to further examine the possibility of an intermediate chiral phase, we show in Figs.\ref{fig3} and \ref{fig4} the temperature ($r^\prime$) dependence of the Binder ratios  and of the correlation-length ratios, respectively. As can be seen from Fig.\ref{fig3}, the Binder ratio of different sizes intersect almost at a common temperature for both cases of the spin and of the chirality at least for $L\leq 40$, whereas, for $L=60$,  the crossing point shows a downshift to lower temperature. This sudden change of the crossing behavior observed at $L=60$ probably reflects the weak first-order nature of the transition observed in the specific heat and the energy distribution of this size. Indeed, both the spin and the chiral Binder ratios of $L=60$ show a deep negative dip, which is a characteristic of a first-order transition. Recall here that just above a first-order transition $T=T_c^+$ the Binder ratio is expected to exhibit a divergent negative dip in the thermodynamic limit. The crossing temperatures of the correlation-length ratios depend on the system sizes only weakly for both cases of the spin and the chirality, as can be seen from Fig.\ref{fig4}.

 Although the difference in the crossing temperatures between different
 physical quantities, either the Binder ratio or the correlation-length
 ratio, or those between different sizes, is sometimes of appreciable amount, we note that the difference in the crossing temperatures between the spin and the chirality  is quite small for a given quantity and given sizes. In Fig.10, we show the size dependence of the spin and the chiral crossing temperatures for both cases of the Binder ratio and the correlation-length ratio. The size is taken here as an average of the two sizes yielding the crossing point, $L_{av}=(L_1+L_2)/2$. As can be seen from the inset, the difference is already very small even for small systems, of order 0.01 which corresponds to $0.1\%$ relative difference, and tends to further decrease with increasing the system size. If we take the crossing temperatures of the Binder ratio between our two largest sizes $L=40$ and $L=60$, the spin and the chiral crossing temperatures, taken here as a measure of the respective transition temperature, are $r_s^\prime = -69.599(2)$ and $r^\prime_c = -69.598(2)$, which coincide within the errors.  The relative difference between the spin and the chiral transition temperature is then limited to $\delta r^\prime < 0.008\%$. Note that this upper limit is significantly smaller than the corresponding estimate obtained from the variational calculation $\delta r^\prime \simeq 42 \%$. Hence, it turns out that the MC results are rather pessimistic about the occurrence of the chiral phase.

 We also try to estimate the bulk transition temperature themselves by
 extrapolating the crossing temperatures to $L=\infty$: See the main
 panel of Fig.10. Such an extrapolation procedure is hampered somewhat
 by the sudden change of the behavior observed at $L=60$. This effect
 seems relatively minor for the correlation-length ratio. By performing a
 power-law fit of the form $r^\prime_{cross}(L) =
 r^\prime_{cross}(L=\infty) + cL^{-\theta}$ to the data of $\xi_s/L$ and
 of $\xi_c/L$, we get $r^\prime_{cross}(L=\infty)=-69.603\pm 0.006$ from
 $\xi_s/L$, and  $r^\prime_{cross}(L=\infty)=-69.600\pm 0.004$ from  $\xi_c/L$.  These estimate are consistent with our estimate based on the Binder ratio given above.
\begin{figure}
 \includegraphics[width=7cm]{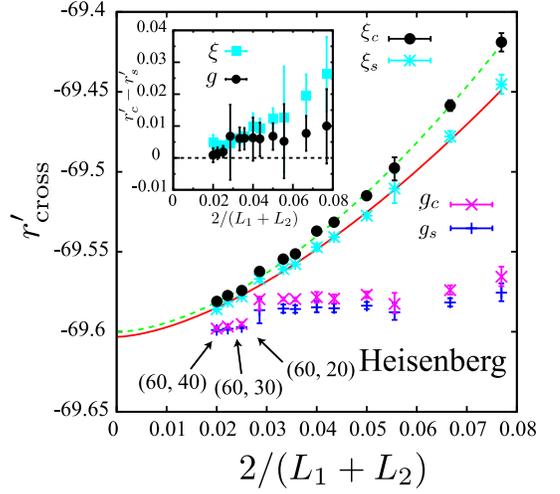}
 \caption{(Color online) The size dependence of the crossing temperatures of the Binder ratio and the correlation-length ratio for both cases of the spin and the chirality of the Heisenberg ($n=3$) chiral GL model. The size is taken here as a mean of the two sizes $L_{av}=(L_1+L_2)/2$, where $L_1$ and $L_2$ are the two liner system sizes yielding a crossing point. The crossing points associated with the $L=60$ data are indicated by arrows. Inset exhibits the size dependence of the  difference between the crossing temperatures of the spin and of the chirality for both cases of the Binder ratio and the correlation-length ratio.}
\label{fig5}
\end{figure}


 We also perform similar MC simulations for other parameter values,
 including $u^\prime=1,10,100,200,1000$ and
 $v^\prime/u^\prime=2.0,3.5$. For all these parameter values, there is
 no evidence of the chiral phase as in the case of $u^\prime=200$ and
 $v^\prime/u^\prime=3.5$ shown above. In case of $u^\prime=1, 10$ and
 $v^\prime/u^\prime=3.5$, signatures of a strong first-order transition
 are observed. By contrast, for other parameter values, the transition
 seems to be continuous, at least in the range of lattice sizes studied here. We summarize our estimates of the spin and the chiral transition temperatures in table \ref{table1}, where each transition temperature is estimated from the crossing temperatures of the Binder ratio between the sizes $L=L_1$ and $L=L_2$. For all cases studied, the difference $\delta r^\prime$ is less than $0.1\%$. 
%
%
\begin{table}
\begin{center}
\begin{tabular}{c|c|c|c|c|c}\hline
 $u^\prime$&$v^\prime/u^\prime$ &$r_s^\prime$ &$r_c^\prime$&$\delta r$&$L1,L2$
 \\ \hline \hline
$ 1$&$2.0$ &$-2.4497(5)$ &$-2.4496(8)$ & $< 0.06\%$ &$16,24$\\
 $10$&$2.0$ &$-17.193(1)$ &$-17.194(1)$&$ < 0.006\%$ & $24,32$\\
 $100$&$2.0$ &$-135.49(2)$ &$-135.48(2)$&$ < 0.04\%$&$24 ,32$\\
 $200$&$2.0$ &$-266.17(4)$ &$-266.12(5)$&$ < 0.06\%$&$16,24$\\ 
 $1000$&$2.0$ &$-1310.9(3)$ &$-1310.7(3)$&$ < 0.07\%$&$16,24$\\
 $100$&$3.5$ &$-36.919(3)$ &$-36.919(3)$&$ < 0.02\%$&$24,32$\\
 $200$&$3.5$ &$-69.599(2)$ &$-69.598(2)$&$ < 0.008\%$&$40,60$\\
 $1000$&$3.5$ &$-330.89(4)$ &$-330.88(2)$&$ < 0.02\%$&$16,24$\\
\hline
\end{tabular}
 \end{center}
 \caption{(Color online) The spin and the chiral transition temperatures and its relative difference  for various parameter values of the Heisenberg ($n=3$) lattice GL Hamiltonian, estimated from the crossing points of the Binder ratio of the two sizes $L=L_1$ and $L_2$.}
\label{table1}
\end{table}

\begin{figure}
 \includegraphics[width=7cm]{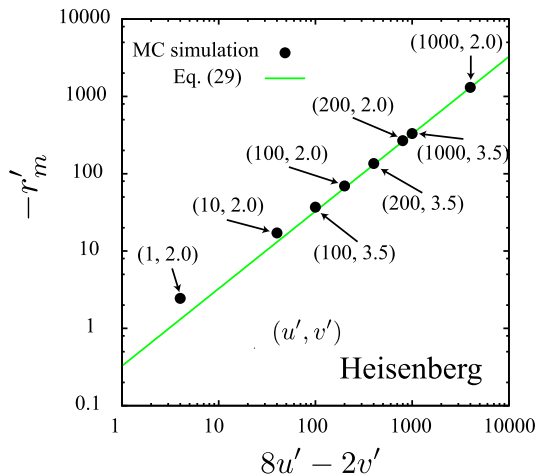}
 \caption{(Color online) The transition temperatures  of the Heisenberg ($n=3$) chiral
 GL model estimated from our Monte Carlo simulations are plotted versus
 the GL Hamiltonian parameter $8u^\prime -2v^\prime$. The line shows the
 relation \eqref{NLS_relation} with $T=3.062$, which is the transition
 temperature of the NL$\sigma$ model reported in ref.\cite{NLS4}
 (Note that we multiply the transition temperature by two due to the
 difference in the definition).} 
 \label{fig6}
\end{figure}

 Finally, we compare the transition temperature of the lattice chiral GL
 model as estimated from our MC with that of the NL$\sigma$ model. In
 Fig.\ref{fig6}, we plot the transition temperature estimated from our
 MC versus the parameter $8u^\prime -2v^\prime$. For $u^\prime > 100$ ,
 the transition temperature can be well fitted by the NL$\sigma$ model
 relation \eqref{NLS_relation}, if one identifies $T$ in
 Eq. \eqref{NLS_relation} as the transition temperature of the lattice
 NL$\sigma$ model reported in ref.\cite{NLS4}. It also indicates that
 the spin and the chirality order simultaneously and there exists only
 single transition in the large $8u^\prime -2v^\prime$ region, since a
 common belief is that there is only single transition in the NL$\sigma$
 model with $R=0$ \cite{NLS,NLS3,NLS2,NLS4}. 
%
%

%
%

\section{Simulation Results: {\it XY} ($n=2$) case\label{XY}}

Next, we present the result of our MC simulation for the {\it XY} case ($n=2$).
As a typical example, we deal with the case of $u^\prime=200$ and 
$v^\prime/u^\prime=3.5$ again. For these parameters, the variational
calculation of section \ref{Analytic} (and appendix \ref{App1}) predicts that the chiral phase is 
stabilized in a wide temperature range between $r^\prime_s \simeq -21.6$ and
$r^\prime_c \simeq -13.4$.  The relative difference between
$r^\prime_c$ and $r^\prime_s$ is $\delta r \simeq 0.47$.

As we discussed in section \ref{Analytic}, the model reduces to the
NL$\sigma$ model in the limit of $u^\prime \to
\infty$. Previous studies showed that, for the case of $n=2$, the NL$\sigma$ model discretized on a 3D lattice exhibited a single first-order transition into the magnetic ordered state \cite{NLS3,NLS_XY}. Meanwhile, the variational calculation predicts the stable chiral phase for sufficiently large $u^\prime$ and $v^\prime/u^\prime$.    

\begin{figure}
 \includegraphics[width=7cm]{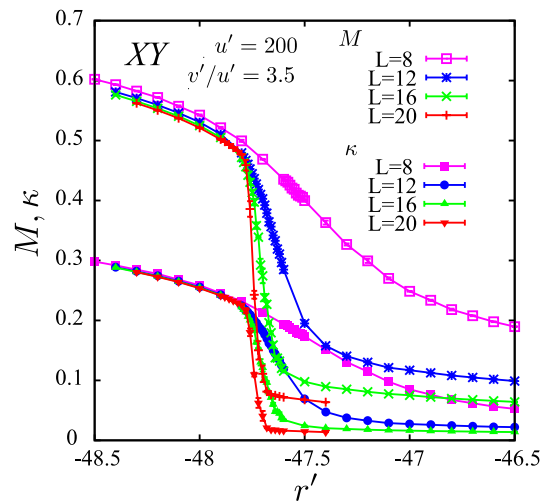}
 \caption{(Color online) The temperature ($r^\prime$) dependence of the spin and the
 chiral order parameters  of the {\it XY\/} ($n=2$) chiral GL model for various sizes. The Hamiltonian parameters
 are set $u^\prime=200$ and $v^\prime/u^\prime=3.5$. }
 \label{fig_xy1}
 \end{figure}

First, we show the temperature ($r^\prime$) dependence of the spin and
the chiral order parameters in Fig.\ref{fig_xy1}. On decreasing the temperature, both order parameters rise up sharply around $r^\prime \simeq 
-47.7$, implying the occurrence of a phase transition around this temperature. Each order parameter rises up at mutually close temperatures such that their relative difference is much smaller than the one predicted from the variational calculation.
The observed onset of the order parameters seems steeper than the one observed in the Heisenberg case. Such a sharp rise of the order parameters is suggestive of a first-order transition.

\begin{figure}
 \includegraphics[width=7cm]{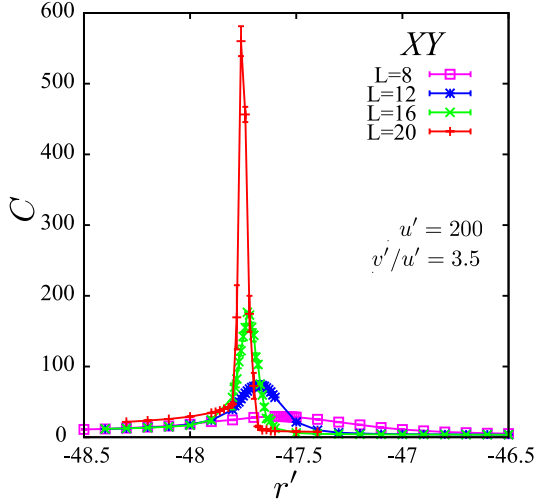}
 \caption{(Color online) The temperature ($r^\prime$) dependence of the specific
 heat of the {\it XY\/} ($n=2$) chiral GL model. The Hamiltonian parameters are set $u^\prime=200$ and
 $v^\prime/u^\prime=3.5$.}
 \label{fig_xy2}
 \end{figure}

\begin{figure}
 \includegraphics[width=7cm]{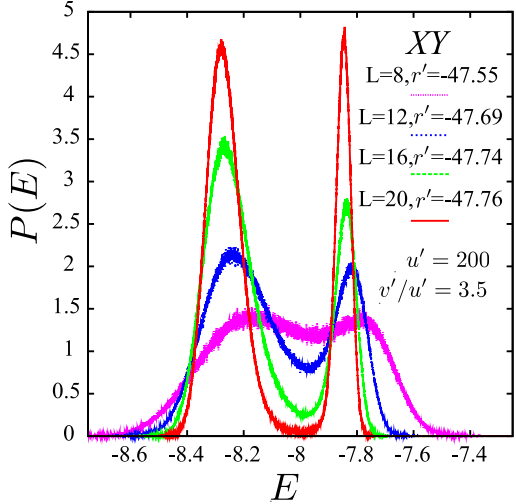}
 \caption{(Color online) The energy distribution near the specific-heat peak temperature  of the {\it XY\/} ($n=2$) chiral GL model for
 various sizes. The Hamiltonian parameters are set $u^\prime=200$ and
 $v^\prime/u^\prime=3.5$.}
 \label{fig_xy2_PE}
\end{figure}

We show  in Fig.\ref{fig_xy2} the temperature ($r^\prime$) dependence of
the specific heat, in Fig.\ref{fig_xy2_PE} the energy distribution near
the transition temperature. The specific heat exhibits only a single
peak with no evidence of successive transitions. Note that the peak
height grows rapidly with increasing the system size $L$, consistently
with a first-order nature of the transition. In fact, the energy
distribution near the transition point shows double peaks characteristic
of a first-order transition. Due to the difficulty in thermalizing the
system exhibiting a rather strong first-order transition, the lattice
sizes in the {\it 
XY} case are restricted to be smaller than those in the Heisenberg case.

 Although we observe a rather strong first-order transition from the
 paramagnetic phase into the helical magnetic phase, there still remains
 a possibility that a chiral phase is stabilized between the tiny
 temperature region between the paramagnetic phase and the helical
 phase. Since the variational calculation of section \ref{Analytic}
 predicted a continuous transition from the paramagnetic phase to the
 chiral phase, we examine here the possibility of a continuous
 transition occurring at a temperature higher than the first-order
 transition temperature. If such a transition really occurs,
 chirality-related dimensionless quantities of various sizes, {\it
 e.g.\/}, the chiral correlation-length ratio and the chiral Binder ratio, are expected to exhibit a crossing behavior at a higher temperature than the first-order transition temperature.

\begin{figure}
 \includegraphics[width=7cm]{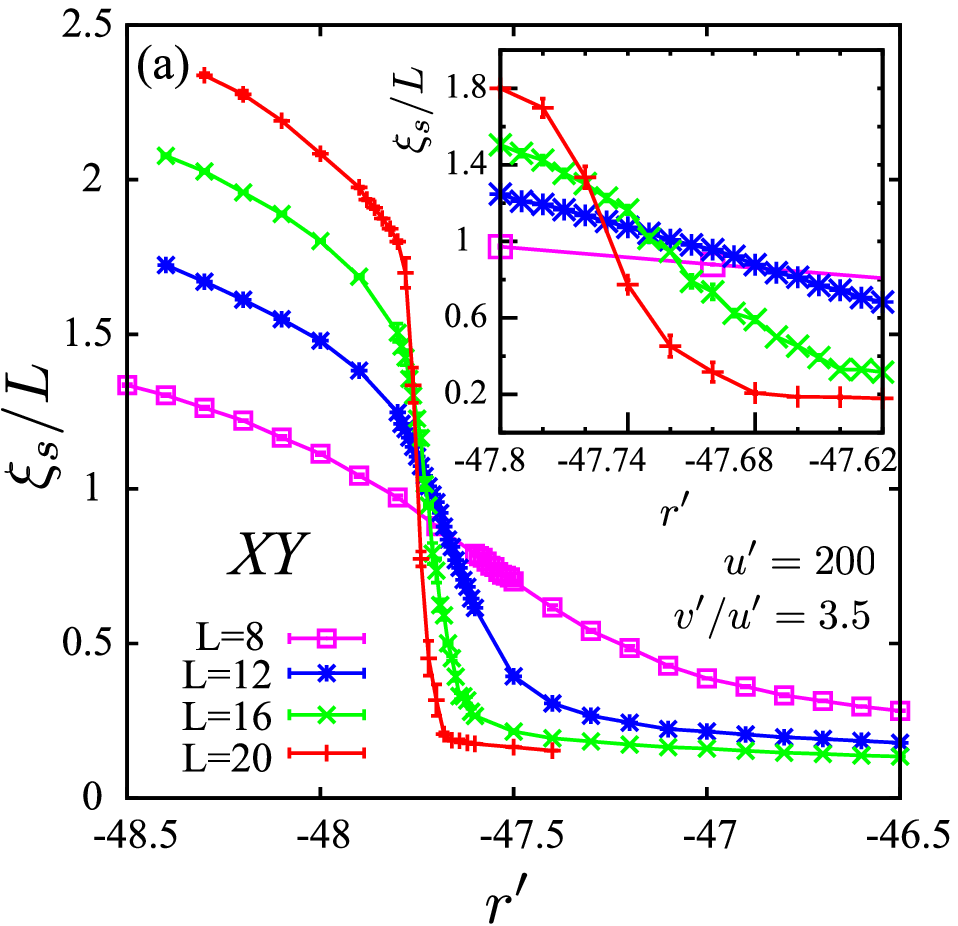}
 \includegraphics[width=7cm]{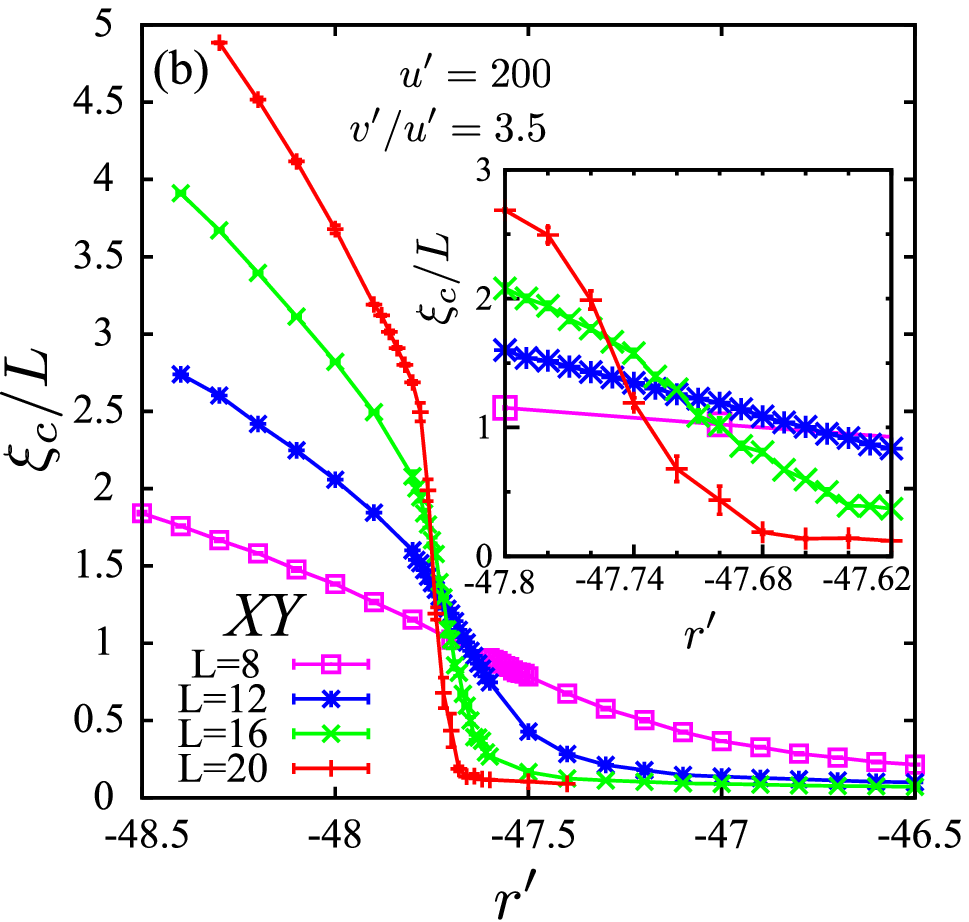}
 \caption{(Color online) The temperature ($r^\prime$) dependence of the correlation
 length ratios for the spin (a), and for the chirality (b),  of the {\it XY\/} ($n=2$) chiral GL model. The
 Hamiltonian parameters are set $u^\prime=200$ and
 $v^\prime/u^\prime=3.5$. Insets are enlarged views of the transition region.}
 \label{fig_xy3}
 \end{figure}

Fig. \ref{fig_xy3} exhibits the temperature ($r^\prime$) dependence of the spin and chiral correlation length ratios. One can see from the figure that, with decreasing the temperature, both $\xi_s/L$ and  $\xi_c/L$ rise up sharply around $r^\prime \simeq -47.75$, close to the specific-heat peak temperature.
The crossing temperatures between our two largest sizes $L=16$ and $L=20$ are $r^\prime_s = -47.758 (5)$ for the spin and $r^\prime_c = -47.753(3)$ for the chirality.  The relative difference between the spin and the chiral crossing temperature is $\delta r^\prime < 0.03\%$. The spin and the chiral crossing points coincide  within the errors, and are close to the first-order transition temperature estimated above.

\begin{figure}
 \includegraphics[width=7cm]{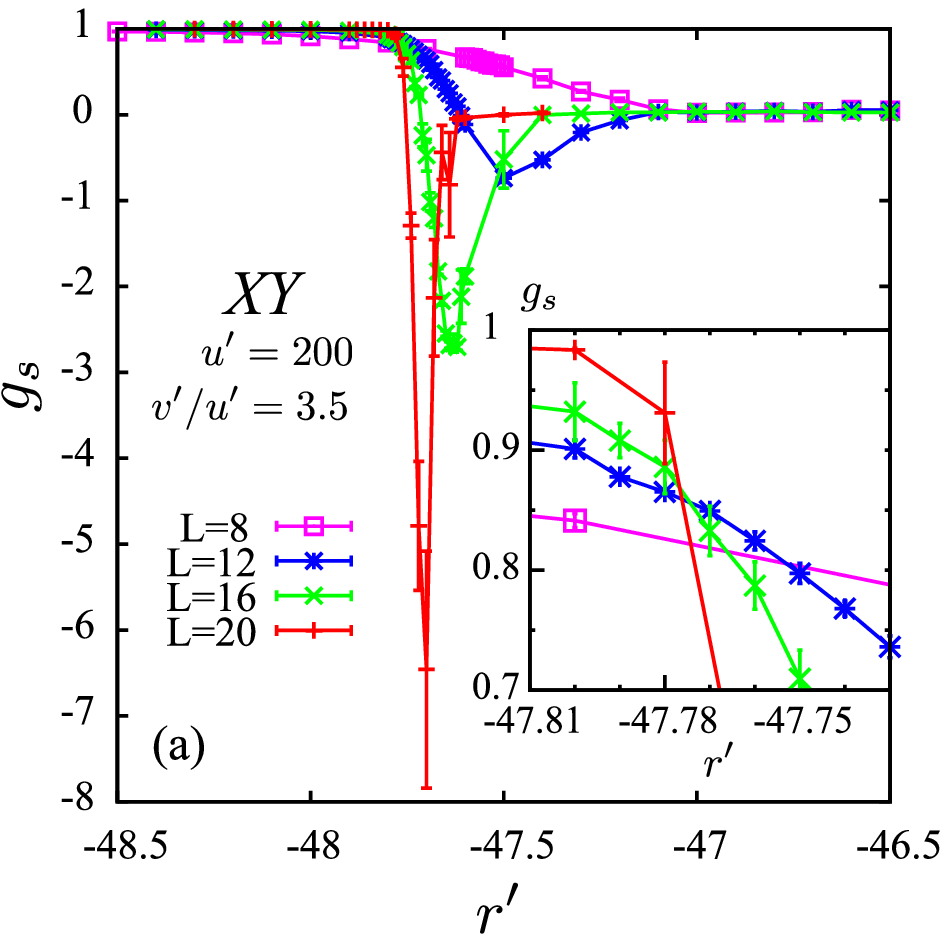}
 \includegraphics[width=7cm]{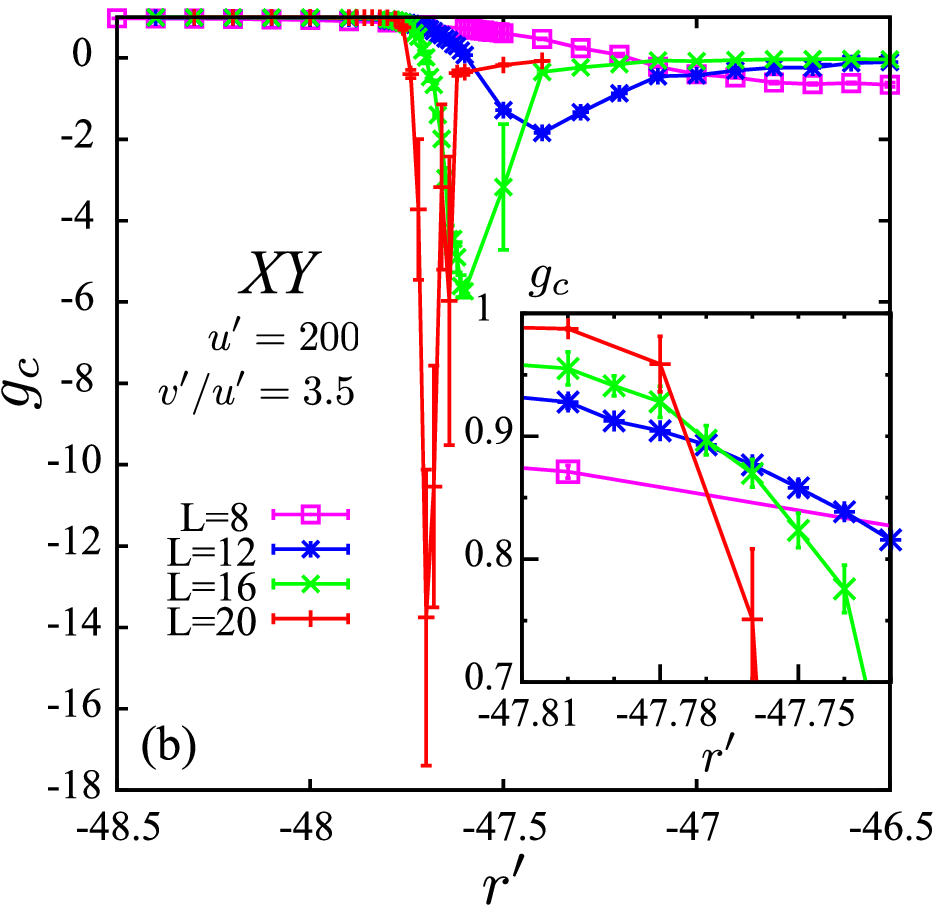}
 \caption{(Color online) The temperature ($r^\prime$) dependence of the Binder ratio for
  the spin (a), and for the chirality (b),  of the {\it XY\/} ($n=2$) chiral GL model. The Hamiltonian parameters
 are set $u^\prime=200$ and $v^\prime/u^\prime=3.5$. Insets are enlarged views of the transition region.}
 \label{fig_xy4}
 \end{figure}

We show in Fig. \ref{fig_xy4} the temperature ($r^\prime$) dependence of
the spin and the chiral Binder ratios. As expected for a first-order
transition, the Binder ratios exhibit a deep negative dip, which grows
with $L$. In addition, the Binder ratios exhibit a crossing on the
positive side of $g$ at a temperature slightly {\it below} the dip
temperature: See the insets.
The observed behavior of the chiral Binder ratio is hardly compatible
with a continuous chiral transition occurring at a temperature higher
than the first-order transition temperature in the thermodynamic
limit. The crossing temperatures 
between our two largest sizes $L=16$ and $L=20$ are $r^\prime_s =
-47.777 (9)$ for the spin and $r^\prime_c = -47.776(8)$ for the chirality. The relative difference between the spin and the chiral transition temperatures is then estimated to be $\delta r^\prime < 0.04\%$

We also perform similar MC simulations for the case of
$v^\prime/u^\prime=2.0$ and $u^\prime=100$, located rather far from the
mean-field tricritical line $v^\prime/u^\prime = 4.0$. Even in this
case, we find a signature of a first-order transition such as the
double-peak structure in the energy distribution. Furthermore, no
evidence of successive transitions are observed. The behaviors of the
correlation-length ratios and the Binder ratios are qualitatively the same as in the case of $v^\prime/u^\prime=3.5$ and $u^\prime=200$, and the relative difference between the spin and the chiral transition temperature is limited to $\delta \tilde{r} <0.05\%$. 



\section{\label{Discussion} Summary and discussion}



In this paper, the nature of the phase transition of regularly
frustrated vector spin systems in three dimension was investigated
based on the chiral GL model both by analytical calculations
and Monte Carlo simulations.  

 We first performed a variational calculation which was an extension of
 the previous calculation by Onoda {\it et al\/}\cite{Onoda}, and showed
 that the 
 chiral phase was stabilized in a certain restricted parameter range. We
 considered  the possibility of a first-order transition, which was not
 considered by Onoda {\it et al\/}. We then found that a first-order
 transition indeed occurred in this model significantly reducing the
 stability range of the chiral phase. Yet, we observed that the chiral
 phase still persisted for sufficiently large $u$ and $v/u$ within the
 variational approximation. We also showed that in the limit of $u \to
 \infty$ the chiral GL model reduced to the NL$\sigma$ model  without the coupling between the chiralities ($R=0$ in eq.(22)).  Previous analysis of the NL$\sigma$ model predicted only a single phase transition directly from the para to the helical phase, which means that there is very little chance for the chiral phase to be stabilized for sufficiently large $u$. 

With reference to these analytic results, we next performed extensive MC simulations on the lattice discretized version of the 3D chiral GL model in search for the possible chiral phase. In contrast to the expectation based on the variational results, however, we did not find any numerical evidence of the chiral phase for both cases of the Heisenberg model ($n=3$) and the {\it XY} model ($n=2$). From the data of the correlation length ratios and the Binder ratios, we conclude that for all cases studied the stability range of the chiral phase, if any, is  less than $0.1\%$ in the relative temperature width.

A possible cause of the appearance of the chiral phase in the
variational calculation might be the fact that, for the chirality,  only
the linear term $\delta a_x\delta b_y-\delta a_y\delta b_x$ is contained
in the variational Hamiltonian \eqref{var_ham}, while the quadratic
terms are also present for the spin. This imbalance inherent to the
variational calculation might lead to an underestimate of chirality
fluctuations compared with spin fluctuations.

Although we could not find any evidence of the chiral phase in the
chiral GL model \eqref{GL_eq}, there still remains a possibility of the
chiral phase originated from some other mechanisms not included in the chiral GL model. If we consider, for example, the direct interaction between the chiralities, $R[\bm{\nabla}(\vec{a}\times\vec{b})]^2 $, it enhances the ordering of
the chirality. For sufficiently large $R$, the chiral phase is trivially stabilized. In fact, David and Jolicoeur showed  on the basis of the NL $\sigma$ model that the chiral phase is
stabilized for sufficiently large $R$ \cite{NLS}. However, for smaller $R$, which is more realistic, their analysis indicated that the chiral phase disappeared and there was a single transition from paramagnetic phase to the helical phase \cite{NLS}. To get further insight into the effect of the $R$-term, we also performed a MC simulation of the chiral GL model with a weak chirality-chirality interaction (the $R$-term) with $u^\prime =200, v^\prime/u^\prime
=2.0$ and $R=0.1$. The results turn out to be qualitatively the same as those of the the original chiral GL model with $R=0$, and $\delta r^\prime$ estimated from the crossing temperature of the Binder ratios is also very small, $\delta r^\prime < 0.02\%$.

 Another mechanism to stabilize the chiral phase in 3D regularly frustrated system was proposed by Villain \cite{Villan_QOD}. He suggested that a chiral phase might be realized in  quasi-one-dimensional {\it XY} spin systems when the inter-chain coupling is sufficiently weak \cite{Villan_QOD}. In purely one-dimensional frustrated {\it XY} spin systems, it has been known that, with decreasing the temperature,  the chirality correlation length increases exponentially, while the spin correlation length diverges as a power law.  By taking into account the effect of the weak inter-chain coupling, Villain predicted that the chiral phase could exist in an intermediate temperature regime above the helical phase.  As far as the authors know, however, a direct numerical evidence of such a chiral phase in quasi-one-dimensional system is still lacking.  Thus, it is an interesting open problem to observe the chiral phase by numerical simulations of such quasi-one-dimensional frustrated spin models.

\begin{acknowledgments}
The authors would like to thank S. Onoda for valuable
 discussions and comments on this work. This work was supported
 by Grant-in-Aid for Scientific Research on Priority Areas ``Novel States
 of Matter Induced by Frustration'' (19052006). We thank 
 the Supercomputer Center, Institute for Solid State Physics,
 University of Tokyo for providing us with the CPU time.
\end{acknowledgments}


\begin{thebibliography}{99}
 \bibitem{Villan} J. Villain, J. Phys. C {\bf 10}, 4793 (1977).
 \bibitem{Kawamura_SG_rev} H. Kawamura, J. Phys. Soc. Jpn. {\bf 79},
	 011007 (2010), and references therein. 
 \bibitem{Kawamura} H. Kawamura Phys. Rev. Lett. {\bf 68}, 3785 (1992).
 \bibitem{Hukushima} K. Hukushima and H. Kawamura, Phys. Rev. B {\bf 72}
        144416 (2005).
 \bibitem{Viet} D.X. Viet and H. Kawamura, Phys. Rev. Lett. {\bf 102}
        027202 (2009); Phys. Rev. B {\bf 80}, 064418 (2009).
 \bibitem{Miyashita} S. Miyashita and H. Shiba, J. Phys. Soc. Jpn. {\bf
        53}, 1145 (1984).
 \bibitem{Ozeki} Y. Ozeki and N. Ito, Phys. Rev. B {\bf 68}, 054414 (2003).
 \bibitem{Hasenbusch} M. Hasenbusch, A. Pelissetto, and E. Vicari,
	 J. Stat. Mech. (2005) P12002.
 \bibitem{Kawamura_XY} H. Kawamura, in {\it Proceeding of the workshop on
	 "Low-Dimensional Quantum Antiferromagnets"} (Fukuoka,
	 Nov. 2001)[arXiv:cond-mat/0202109], and references therein.
 \bibitem{Onoda} S. Onoda and N. Nagaosa, Phys. Rev. Lett. {\bf 99},
       027206 (2007).

 \bibitem{Kawamura_rev} H. Kawamura , J. Phys.:Condens. Matter  {\bf
        10}, 4707 (1998), and references therein.
 \bibitem{Kawamura_1992} H. Kawamura, J. Phys. Soc. Jpn. {\bf 61} 1299 (1992).
 \bibitem{MC_stack2} A. Mailhot, M. L. Plumer and A. Caill\'{e},
        Phys. Rev. B {\bf 50} 6854 (1994).
 \bibitem{MC_stack3} E. H. Boubcherur, D. Loison and H. T. Diep,
        PHys. Rev. B {\bf 54}, 4165 (1996).
 \bibitem{MC_first} M. L. Plumer and A. Mailhot, J. Phys.:Condens. Matt. {\bf
        9}, L165  (1997).
 \bibitem{MC_stack4} V. Thanh Ngo and H. T. Diep, J. Appl. Phys. {\bf
        103}, 07C712  (2008).
       
 \bibitem{Kawamura_1988} H. Kawamura , Phys. Rev. B {\bf 38}, 4916 (1988).
 \bibitem{Vicari1} A. Pelissetto, P. Rossi, and E. Vicari, Phys. Rev. B
	 {\bf 63}, 140414(R) (2001). 
 \bibitem{Vicari2} P. Calabrese, P. Parruccini, A. Pelissetto, and
	 E. Vicari, Phys. Rev. B {\bf 70}, 174439 (2004). 
 \bibitem{Exp1} T. E. Mason, M. F. Collins and B.D. Gaulin, J. Phys. C:
        Solid State Phys. {\bf 20}, L945 (1987), T. E. Mason,
        B.D. Gaulin and M. F. Collins , Phys. Rev. B {\bf 39}, 586 (1989).
 \bibitem{Exp2} Y. Ajiro, T. Nakashima, Y. Unno, H. Kadowaki, M. Mekata
        and N. Achiwa, J. Phys. Soc. Jpn. {\bf 57} 2648 (1988),
        H. Kadowaki, S. M. Shiapiro, T. Inami and Y. Ajiro,
        J. Phys. Soc. Jpn. {\bf 57} 2640 (1988).
 \bibitem{Exp3} K. Takeda, N. Ury\^{u}, K. Ubukoshi and
        K. Hirakawa, J. Phys. Soc. Jpn. {\bf 55} 727 (1986).
 \bibitem{Exp4} H. Kadowaki, K. Ubukoshi and K. Hirakawa,
        J. Phys. Soc. Jpn. {\bf 56} 4027 (1987).       
 \bibitem{Exp5} J. Wosnitza, R. Deutschmann, H. von L\"{o}hneysen and
        R. K. Kremer J. Phys.:Condens. Matt. {\bf
        6}, 8045  (1994).
 \bibitem{stack_first} M. Itakura J.Phys. Soc. Jpn. {\bf
        72}, 74  (2003).
 \bibitem{NLS} F. David and  T. Jolicoeur, Phys. Rev. Lett. {\bf 76},
       3148 (1996).
 \bibitem{NLS3} H. Kunz and G. Zumbach , J. Phys. A: Math. Gen. {\bf 26},
       3121 (1993).
 \bibitem{NLS2} H. T. Diep and D. Loison , J. Appl. Phys. {\bf 76},
       15 (1994).
 \bibitem{NLS4} D. Loison and K. D. Schotte, Eur. Phys. J. B {\bf 14},
       125 (2000).
 \bibitem{NLS_XY} D. Loison and K. D. Schotte , Eur. Phys. J. B {\bf 5},
       735 (1998).
       
\bibitem{Villan_QOD} J. Villain, in {\it proceedings of the 13th IUPAP
       Conference on Statistical Physics} [Ann. Isr. Phys. Soc. 2, 565 (1978)].
\end{thebibliography}

\appendix
\section{Details of a variational approximation\label{App1}}
\subsection{Free energy}

In this Appendix, we explicitly show the form of the trial free energy $\mathcal{F}_0 + \langle \mathcal{H} -\mathcal{H}_0\rangle_0$.

Since the variational Hamiltonian (6) is diagonalized with respected to
$\vec{\alpha}$, $\vec{\beta}$ and $\vec{\gamma}$ as given in (8), the term 
$\mathcal{F}_0$ is easily calculated as
\begin{align}
 \mathcal{F}_0&\equiv-\log\left\{\int
 \prod_{\bm{q}}d\delta\vec{a}_{\bm{q}}d\delta\vec{b}_{\bm{q}}
 \exp(-\mathcal{H}_0)\right\} \notag \\
&=
 -\sum_{\bm{q}}\Biggl\{\log\left[\frac{\pi}{V(q^2+r_\parallel-h_\kappa/2)}\right]\notag
 \\
 &\qquad +\log\left[\frac{\pi}{V(q^2+r_\parallel+h_\kappa/2)}\right]
 + \log\left[\frac{\pi}{V(q^2+r_\perp)}\right]\Biggr\}.
\label{eq_F0}
\end{align}

In calculating the second term $\langle \mathcal{H}
-\mathcal{H}_0\rangle_0$, we rewrite eq.(1) in terms of
$\alpha$, $\beta$ and $\gamma$ defined by eq.(7), to get

\begin{widetext}
 \begin{eqnarray}
 \mathcal{H} = \frac{1}{2}\int
 d\bm{x}\Biggl\{r\left(\vec{\alpha}^2+\vec{\beta}^2+\vec{\gamma}^2\right) +
         (\bm{\nabla}\vec{\alpha})^2 + (\bm{\nabla}\vec{\beta})^2
 +(\bm{\nabla}\vec{\gamma})^2 
 +\left(u-\frac{1}{4}v\right)\left(\vec{\alpha}^4+\vec{\beta}^4\right)
        +
 2\left(u+\frac{1}{4}v\right)\vec{\alpha}^2\vec{\beta}^2 \nonumber\\ 
+ u\vec{\gamma}^4
 +2\left(u-\frac{1}{4}v\right)\vec{\gamma}^2
 \left(\vec{\alpha}^2+\vec{\beta}^2\right) 
 +v\Bigl[2\gamma_1\gamma_2(\vec{\alpha}\cdot\vec{\beta})+(\gamma_1^2
 -\gamma_2^2)(\alpha_2\beta_1-\alpha_1\beta_2)\Bigr]
\Biggr\}.
\end{eqnarray}
\end{widetext}

By substituting $\vec{\alpha}=\delta\vec{\alpha}+\vec{A}$, {\it etc\/},
into (A.2), $\langle \mathcal{H}-\mathcal{H}_0\rangle_0$ is calculated via simple  Gaussian integrals as 

\begin{widetext}
\begin{eqnarray}
 \langle \mathcal{H}-\mathcal{H}_0\rangle_0=
 \frac{V}{2}\Biggl\{\Biggl[r-\left(r_\parallel-\frac{h_\kappa}{2}\right)+4\left(u-\frac{v}{4}\right)\vec{A}^2 +2\left(u+\frac{v}{4}\right)\vec{B}^2+2\left(u-\frac{v}{4}\right)\vec{C}^2\Biggr]
 \sum_{\bm{q}}\langle\delta\vec{\alpha}_{\bm{q}}\cdot
 \delta\vec{\alpha}_{-\bm{q}}\rangle_0  \nonumber \\
 +\Biggl[r-\left(r_\parallel+\frac{h_\kappa}{2}\right)+2\left(u+\frac{v}{4}\right)\vec{A}^2
 +4\left(u-\frac{v}{4}\right)\vec{B}^2 +2\left(u-\frac{v}{4}\right)\vec{C}^2\Biggr]\sum_{\bm{q}}\langle\delta\vec{\beta}_{\bm{q}}\cdot
 \delta\vec{\beta}_{-\bm{q}}\rangle_0 \nonumber \\
 +\left[r-r_\perp+2\left(u-\frac{v}{4}\right)\left(\vec{A}^2+\vec{B}^2\right)
 +4u\vec{C}^2\right]
 \sum_{\bm{q}}\langle\delta\vec{\gamma}_{\bm{q}}\cdot
 \delta\vec{\gamma}_{-\bm{q}}\rangle_0  \nonumber\\
+2\left(u-\frac{v}{4}\right)
  \left[\left(\sum_{\bm{q}}\langle\delta\vec{\alpha}_{\bm{q}}\cdot
 \delta\vec{\alpha}_{-\bm{q}}\rangle_0\right)^2
 +\left(\sum_{\bm{q}}\langle\delta\vec{\beta}_{\bm{q}}\cdot
 \delta\vec{\beta}_{-\bm{q}}\rangle_0\right)^2\right]  \nonumber\\
 +2\left(u+\frac{v}{4}\right)
 \left(\sum_{\bm{q}}\langle\delta\vec{\alpha}_{\bm{q}}\cdot
 \delta\vec{\alpha}_{-\bm{q}}\rangle_0\right) 
 \left(\sum_{\bm{q}}\langle\delta\vec{\beta}_{\bm{q}}\cdot
 \delta\vec{\beta}_{-\bm{q}}\rangle_0\right) +2u\left(\sum_{\bm{q}}\langle\delta\vec{\gamma}_{\bm{q}}\cdot
 \delta\vec{\gamma}_{-\bm{q}}\rangle_0\right)^2\nonumber\\
 +2\left(u-\frac{v}{4}\right)\left(\sum_{\bm{q}}\langle\delta\vec{\gamma}_{\bm{q}}\cdot
 \delta\vec{\gamma}_{-\bm{q}}\rangle_0\right)
\left[\left(\sum_{\bm{q}}\langle\delta\vec{\alpha}_{\bm{q}}\cdot
 \delta\vec{\alpha}_{-\bm{q}}\rangle_0\right)
 +\left(\sum_{\bm{q}}\langle\delta\vec{\beta}_{\bm{q}}\cdot
 \delta\vec{\beta}_{-\bm{q}}\rangle_0\right)\right]\nonumber\\
+r(\vec{A}^2+\vec{B}^2+\vec{C}^2)+\left(u-\frac{1}{4}v\right)(\vec{A}^4
 + \vec{B}^4)
 +2\left(u+\frac{1}{4}v\right)\vec{A}^2\vec{B}^2 +u\vec{C}^4 \nonumber\\
 + 2\left(u-\frac{1}{4}v\right)\vec{C}^2(\vec{A}^2+\vec{B}^2) 
 + v\biggl[2C_1C_2(\vec{A}\cdot\vec{B}) +(C_1^2-C_2^2)(A_2B_1-A_1B_2)\biggr]\Biggr\}.
\end{eqnarray}
\end{widetext}

In the case of $A_1=m, A_2=0$, $\vec{B}=\vec{0}$, $\vec{C}=\vec{0}$, as assumed  in \S IIIA, the trial free energy is simplified as

\begin{widetext}
  \begin{eqnarray}
 \mathcal{F}_0 +\langle\mathcal{H}-\mathcal{H}_0\rangle_0 =
  -\frac{V\Lambda^3}{2\pi^2}\int_0^1 dq
  \Biggl\{\log\left[\frac{1}{
q^2+\tilde{r}_\parallel-\tilde{h}_\kappa/2}\right]  
 +\log\left[\frac{1}{q^2+\tilde{r}_\parallel+\tilde{h}_\kappa/2}\right]
 + \log\left[\frac{1}{q^2+\tilde{r}_\perp}\right]\Biggr\} \nonumber\\
+\frac{V\Lambda^3}{4\pi^2}
 \Biggl\{\left[\tilde{r}-\tilde{r}_\parallel+\frac{\tilde{h}_\kappa}{2}+4\left(\tilde{u}-\frac{\tilde{v}}{4}\right)\tilde{m}^2\right]\sigma^2_\alpha 
+\left[\tilde{r}-\tilde{r}_\parallel-\frac{\tilde{h}_\kappa}{2}+2\left(\tilde{u}+\frac{\tilde{v}}{4}\right)\tilde{m}^2
\right]\sigma^2_\beta \nonumber \\
 +\left[\tilde{r}-\tilde{r}_\perp+2\left(\tilde{u}-\frac{\tilde{v}}{4}\right)\tilde{m}^2\right]\sigma^2_\gamma 
+2\left(\tilde{u}-\frac{\tilde{v}}{4}\right)\left(\sigma_\alpha^4+\sigma_\beta^4\right)
 +2\left(\tilde{u}+\frac{\tilde{v}}{4}\right) \sigma^2_\alpha\sigma^2_\beta\notag\\
 +2\tilde{u}\sigma_\gamma^2+2\left(\tilde{u}-\frac{\tilde{v}}{4}\right)\sigma^2_\gamma
\left(\sigma^2_\alpha+\sigma^2_\beta\right)
 +\tilde{r}\tilde{m}^2+\left(\tilde{u}-\frac{1}{4}\tilde{v}\right)\tilde{m}^4\Biggr\}
  + \text{const.},
\end{eqnarray}
\end{widetext}

where we introduced the cutoff wavevector $\Lambda$, and scaled various parameters as in \eqref{scaled_var}. By taking the derivatives of $\mathcal{F}_0+\langle\mathcal{H}-\mathcal{H}_0\rangle_0$ with respect to $\tilde{r}_\parallel$, $\tilde{r}_\perp$, $\tilde{h}_\kappa$ and 
$\tilde{m}$, and setting them to zero, we can get the conditions for
the optimal parameter values given in \eqref{eq_r0}-\eqref{eq_m}.

\subsection{\label{A2}Relation between $\tilde{r}_s$ and $\tilde{r}_c$}

In this Appendix, the relation between the para-to-chiral
 and the chiral-to-helical  continuous transition temperatures, $\tilde{r}_c$ and $\tilde{r}_s$, is investigated. We prove here the existence of a
critical value $\tilde{u}_c$ such that $\tilde{r}_s > \tilde{r}_c$ for
$\tilde{u} < \tilde{u}_c$ and $\tilde{r}_s < \tilde{r}_c$ for $\tilde{u} >
\tilde{u}_c$.

 To simplify the notation, we define a function

\begin{align}
 g(r) &\equiv \int_0^1 \frac{q^2}{q^2+r} dq \notag \\
 &=1 -\sqrt{r}\mathrm{atan}\frac{1}{\sqrt{r}}.
\end{align}

By using $g(r)$, the variances of $\tilde{\alpha}$, $\tilde{\beta}$
and $\tilde{\gamma}$ are calculated as 

\begin{align}
 \sigma^2_\alpha&=2g(\tilde{r}_\parallel-\tilde{h}_\kappa/2), \notag \\
 \sigma^2_\beta&=2g(\tilde{r}_\parallel+\tilde{h}_\kappa/2), \notag \\
 \sigma^2_\gamma&=2g(\tilde{r}_\perp).
\end{align}

The conditions of optimal parameter values
eqs.\eqref{eq_r0}-\eqref{eq_k}  
may be given with $\tilde{m}=0$ by

\begin{align}
  \tilde{r}_\parallel &= \tilde{r} +
  2\left(3\tilde{u}-\frac{\tilde{v}}{4}\right)\left[
 g(\tilde{r}_\parallel-\tilde{h}_\kappa/2)+
 g(\tilde{r}_\parallel+\tilde{h}_\kappa/2)\right] \notag \\
 &\qquad+4\left(\tilde{u}-\frac{\tilde{v}}{4}\right)g(\tilde{r}_\perp),
\label{opt_eq_g1}
\\
\tilde{r}_\perp&=r+4\left(\tilde{u}-\frac{\tilde{v}}{4}\right)\left[
 g(\tilde{r}_\parallel-\tilde{h}_\kappa/2)+
 g(\tilde{r}_\parallel+\tilde{h}_\kappa/2) \right] \notag \\
&\qquad+ 8\tilde{u}g(\tilde{r}_\perp) \label{opt_eq_g2}, 
\\
 1&=
 \left(3\tilde{v}-4\tilde{u}\right)\left(g(\tilde{r}_\parallel-\tilde{h}_\kappa/2)-
 g(\tilde{r}_\parallel+\tilde{h}_\kappa/2)\right)/\tilde{h}_\kappa.
\label{opt_eq_g3}
\end{align} 

The equations determining the critical parameters $\tilde{r}^{(c)}_\parallel$
and $\tilde{r}^{(s)}_\parallel$, eqs.(20) and (21), 
are also given by 

\begin{align}
 (3\tilde{v}-4\tilde{u})f(\tilde{r}^{(c)}_\parallel) &=1, \label{eq_rc}\\
 (3\tilde{v}-4\tilde{u})\left[g(0)-g(2\tilde{r}^{(s)}_\parallel)\right] &=2\tilde{r}^{(s)}_\parallel,
\end{align}
where the function $f(r)$ is defined by

\begin{align}
 f(r)&\equiv -\frac{d g}{d r}\notag\\
 &= \frac{1}{2}\left[\frac{1}{\sqrt{r}}\mathrm{atan}\frac{1}{\sqrt{r}} -\frac{1}{1+r}\right].
\end{align}

Originally, eqs.\eqref{opt_eq_g1} - \eqref{opt_eq_g3} are the equations determining the variational parameters $\tilde{r}_\parallel$, $\tilde{r}_\perp$ and $\tilde{h}_\kappa$ as functions of the Hamiltonian parameters $\tilde{r}$, $\tilde{u}$ and $\tilde{v}$. One can also regard $\tilde{r}$ given as a function of $\tilde{h}_\kappa$, $\tilde{u}$ and $\tilde{v}$. In order to discuss the relation between $\tilde{r}_s$ and $\tilde{r}_c$, we examine here the behavior of $\tilde{r}(\tilde{h}_\kappa; \tilde{u}, \tilde{v})$ around  $\tilde{h}_\kappa=0$ for given $\tilde{u}$ and $\tilde{v}$. Because of symmetry, $d \tilde{r}/ d \tilde{h}_\kappa$ is equal to zero at $\tilde{h}_\kappa=0$ (at $\tilde{r}=\tilde{r}_c$). Hence, the sign of the second derivative of $\tilde{r}(\tilde{h}_\kappa)$ at the chiral transition point $\tilde{h}_\kappa \rightarrow 0$ 
determines the relation between $\tilde{r}_s$ and $\tilde{r}_c$. Namely, $\tilde{r}_s < \tilde{r}_c$ if $d^2\tilde{r}/d \tilde{h}_\kappa^2 < 0$, and $\tilde{r}_s > \tilde{r}_c$ if $d^2\tilde{r}/d \tilde{h}_\kappa^2 > 0$. 

By taking the second derivatives of equations \eqref{opt_eq_g1} - \eqref{opt_eq_g3} with  respect to $\tilde{h}_\kappa$ and setting $\tilde{r}_\parallel=\tilde{r}^{(c)}_\parallel$,  $\tilde{r}_\perp=\tilde{r}^{(c)}_\parallel$ and $\tilde{h}_\kappa = 0$, the second derivative  $d^2\tilde{r}/d \tilde{h}_\kappa^2$ is given by

\begin{multline}
 \frac{d^2\tilde{r}}{d
 \tilde{h}_\kappa^2} = \left[1 +
		      \left(16\tilde{u}-2\tilde{v}\right)f(\tilde{r}^{(c)}_\parallel)\right] 
 \left[-\frac{f^{\prime\prime}(\tilde{r}^{(c)}_\parallel)}{12 f^\prime(\tilde{r}^{(c)}_\parallel)}\right] \\
+
 \frac{3\tilde{u}-\frac{\tilde{v}}{4}+
  \left(4\tilde{u}-\frac{\tilde{v}}{2}\right)
  \left(4\tilde{u}+\tilde{v}\right)
  f(\tilde{r}^{(c)}_\parallel)}{1+\left(4\tilde{u}+\tilde{v}\right)f(\tilde{r}^{(c)}_\parallel)   
  }f^\prime(\tilde{r}^{(c)}_\parallel),
\label{eq_balance}
\end{multline}
where $f^\prime(r)$ and $f^{\prime\prime}(r)$ are the first and the
second derivatives of $f(r)$, respectively. By use of eq.\eqref{eq_rc},
eq.\eqref{eq_balance} can be rewritten as 

\begin{equation}
 \label{drdk}
  \frac{d^2\tilde{r}}{d\tilde{h}_\kappa^2} =
   \left\{\frac{C_1(s)}{C_2(s)}-\frac{\left[f^\prime(\tilde{r}^{(c)}_\parallel)\right]^2} 
   {f(\tilde{r}^{(c)}_\parallel)f^{\prime\prime}(\tilde{r}^{(c)}_\parallel)}\right\}C_2(s)\left[-\frac{f^{\prime\prime}(\tilde{r}^{(c)}_\parallel)}{f^\prime(\tilde{r}^{(c)}_\parallel)}\right], 
\end{equation}
where we set $s\equiv \tilde{v}/\tilde{u}$, while the functions $C_1(s)$ and $C_2(s)$ are defined by

\begin{align}
 C_1(s)&\equiv \frac{1}{12}\left[1 + \frac{16-2s}{3s-4}\right], \\
 C_2(s)&\equiv \frac{3-\dfrac{s}{4} + \left(4
 -\dfrac{s}{2}\right)\dfrac{4+s}{3s-4}}{4s}.
\end{align}

Note that $C_1(s)$ and $C_2(s)$ are positive for $4/3 < s < 4 $. It is easily confirmed that $C_1(s)/C_2(s)$ is an increasing function of $s$, lying in the range  $1/3 < C_1(s)/C_2(s) < 2/3$ for $4/3 < s < 4$. One can also show that the function $[f^\prime(r)]^2/[f(r)f^{\prime\prime}(r)]$ is an increasing function of $r$ in the range between $1/3$ and $2/3$ for $r>0$. Hence, there exists a critical value $r^\ast(s)$ such that $d^2 \tilde{r}/d \tilde{h}_\kappa^2 > 0$ for $\tilde{r}_\parallel^{(c)} < r^\ast(s)$, and $d^2 \tilde{r}/d\tilde{h}_\kappa^2 < 0$ for $\tilde{r}_\parallel^{(c)} > r^\ast(s)$, where $r^\ast(s)$ is given by the solution of the equation

\begin{equation}
\frac{C_1(s)}{C_2(s)}-\frac{\left[f^\prime(r^\ast)\right]^2}{f(r^\ast)f^{\prime\prime}(r^\ast)} 
 = 0.
\end{equation}

We have $\tilde{r}_c > \tilde{r}_s$ if and only if $\tilde{r}^{(c)}_\parallel > r^\ast(s)$, while $\tilde{r}_c < \tilde{r}_s$ if $\tilde{r}^{(c)}_\parallel < r^\ast(s)$.

From eq.\eqref{eq_rc}, $\tilde{r}_\parallel^{(c)}$ is an increasing function of $\tilde{u}$ for fixed $s=\tilde{v}/\tilde{u}$, the critical value $\tilde{u}_c(s)$ corresponding to $r^\ast$ is given by

\begin{equation}
 \tilde{u}_c(s) \equiv \frac{1}{(3s-4)f(r^\ast)}.
\end{equation}

Thus, the same conclusion can be restated in terms of $\tilde{u}$ as $\tilde{r}_c > \tilde{r}_s$ if and only if $\tilde{u} > \tilde{u}_c(s)$. This
critical value $\tilde{u}_c(s)$ is an increasing function of $s=\tilde{v}/\tilde{u}$ and is in the range $0 < \tilde{u}_c < \infty$ for $ 4/3 < \tilde{v}/\tilde{u} < 4$.

\subsection{Variational approximation for the {\it XY} spin ($n=2$)}

In this Appendix, we show the results of the variational
calculation for the {\it XY} case ($n=2$). As
mentioned in section \ref{Analytic}, the 
variational Hamiltonian for the {\it XY} case may be given by

\begin{multline}
 \mathcal{H}_0 = \frac{V}{2}\sum_{\bm{q}}
  \Bigl[\left(q^2+r_\parallel-\frac{h_\kappa}{2}\right)
 \delta\vec{\alpha}_{\bm{q}}\cdot  
   \delta\vec{\alpha}_{\bm{-q}} \\
   +\left(q^2+r_\parallel+\frac{h_\kappa}{2}\right)
 \delta\vec{\beta}_{\bm{q}}\cdot  
   \delta\vec{\beta}_{\bm{-q}}.
\end{multline}

As in the Heisenberg case, we assume that $A_1 = m \ge 0$,
$A_2=0$ and $\vec{B}=\vec{0}$. The trial free energy 
$\mathcal{F}_0+\left\langle\mathcal{H}-\mathcal{H}_0\right\rangle_0$ is
then calculated as  

\begin{widetext}
 \begin{eqnarray}
 \mathcal{F}_0 +\left\langle\mathcal{H}-\mathcal{H}_0\right\rangle_0 = -\frac{V\Lambda^3}{2\pi^2}\int_0^1 dq  \Biggl\{\log\left[\frac{1}{q^2+\tilde{r}_\parallel-\tilde{h}_\kappa/2}\right] 
  +\log\left[\frac{1}{q^2+\tilde{r}_\parallel+\tilde{h}_\kappa/2}\right]
 \nonumber \\
+\frac{V\Lambda^3}{4\pi^2}\Biggl\{\left[\tilde{r}-\tilde{r}_\parallel+\frac{\tilde{h}_\kappa}{2}
 +4\left(\tilde{u}-\frac{\tilde{v}}{4}\right)\tilde{m}^2\right]\sigma^2_\alpha
 +\left[\tilde{r}-\tilde{r}_\parallel-\frac{\tilde{h}_\kappa}{2}+2\left(\tilde{u}+\frac{\tilde{v}}{4}\right)\tilde{m}^2 
\right]\sigma^2_\beta \nonumber \\
 +2\left(\tilde{u}-\frac{\tilde{v}}{4}\right)\left(\sigma_\alpha^4+\sigma_\beta^4\right) 
 +2\left(\tilde{u}+\frac{\tilde{v}}{4}\right)
 \sigma^2_\alpha\sigma^2_\beta
+\tilde{r}\tilde{m}^2+\left(\tilde{u}-\frac{1}{4}\tilde{v}\right)\tilde{m}^4\Biggr\}
 + \text{const.}
\end{eqnarray}
\end{widetext}

By taking the derivatives of $\mathcal{F}_0
+\langle\mathcal{H}-\mathcal{H}_0\rangle_0$ with respected to
$\tilde{r}_\parallel$, $\tilde{h}_\kappa$ and $\tilde{m}$, and setting them
to zero, we get the following conditions for the optimal
parameters values,

\begin{align}
 \tilde{r}_\parallel &= \tilde{r} +
  \left(3\tilde{u}-\frac{\tilde{v}}{4}\right)\left(\tilde{m}^2+\sigma^2_\alpha+\sigma^2_\beta\right), \label{eq_r0_xy}\\
 \tilde{h}_\kappa&=
 2\left(\frac{3}{4}\tilde{v}-\tilde{u}\right)\left(\tilde{m}^2+\sigma^2_\alpha-\sigma^2_\beta\right), 
  \label{eq_k_xy} \\
0&=\Biggl[
  2\left(\tilde{u}-\frac{\tilde{v}}{4}\right)\tilde{m}^2- \left(\tilde{r}_\parallel-\frac{\tilde{h}_\kappa}{2}\right)\Biggr]\tilde{m} \label{eq_m_xy}.
\end{align}

The difference between these equations and the corresponding equations of
the Heisenberg case \eqref{eq_r0}-\eqref{eq_m} is only the terms related to
$\tilde{r}_\perp$ or $\sigma_\gamma^2$ which are simply absent in the {\it XY\/} case. Especially, eq.\eqref{eq_k_xy} is exactly the same as eq.\eqref{eq_k}. Therefore, when $\tilde{v}/{u} > 4/3$, there is a continuous transition
to the chiral phase at the same critical value of $\tilde{r}_\parallel=
\tilde{r}_\parallel^{(c)}$ as in the Heisenberg case. The
continuous transition between the chiral phase and the helical
phase also occurs at the same critical value $\tilde{r}_\parallel=
\tilde{r}_\parallel^{(s)}$ as in the Heisenberg case.
On the other hand, if one analyzes how the ordering proceeds when the
temperature $\tilde{r}$ is varied,  
the transition temperatures of the {\it XY} spins differ  from
those of the Heisenberg model quantitatively. Both $\tilde{r}_s$ and $\tilde{r}_c$ are generally higher than the corresponding transition temperatures of 
the Heisenberg model.

By use of the same technique as in Appendix \ref{A2}, one can show that there is the critical value $\tilde{u}_c$ such that $\tilde{r}_c > \tilde{r}_s$ for $\tilde{u} > \tilde{u}_c $. From eqs.\eqref{eq_r0_xy} and \eqref{eq_k_xy}, the same form of equation for $d \tilde{r}^2/d \tilde{h}_\kappa^2$ with \eqref{drdk} can be derived, where the functions $C_1(s)$ and $C_2(s)$ are different from those of the
Heisenberg case and are given by

\begin{align}
C_1(s) &\equiv \frac{1}{12}\left[1 + \frac{12-s}{3s-4}\right],\\
 C_2(s) &\equiv \frac{3-\dfrac{s}{4}}{3s-4}.
\end{align}

Again, $C_1(s)$ and $C_2(s)$ are positive for $4/3 < s <4$, while the ratio $C_1(s)/C_2(s)$ has the same characteristics as those of 
the Heisenberg case such that it is an increasing function of $s$ satisfying $1/3 < C_1(s)/C_2(s) < 2/3$. Therefore, we reach the same
conclusion as in the Heisenberg case, {\it i.e.}, $\tilde{r}_c
>\tilde{r}_s$, if and only if $\tilde{u} > \tilde{u}_c(s)$. 
Note that the value of $\tilde{u}_c(s)$ differs from that of the
Heisenberg model due to the difference in the function $C_1(s)/C_2(s)$.

\begin{figure*}
 \includegraphics[width=7cm]{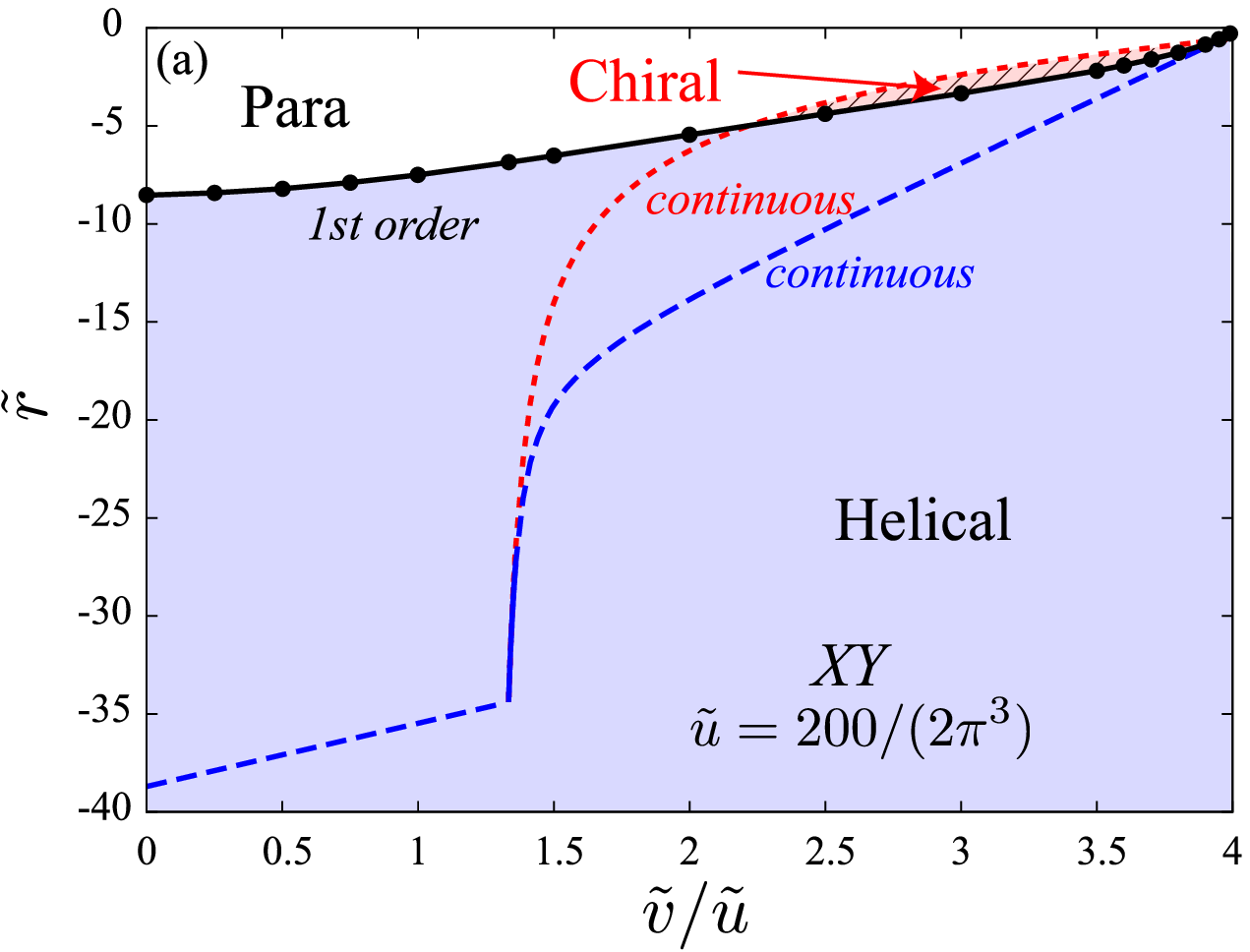}
 \includegraphics[width=7cm]{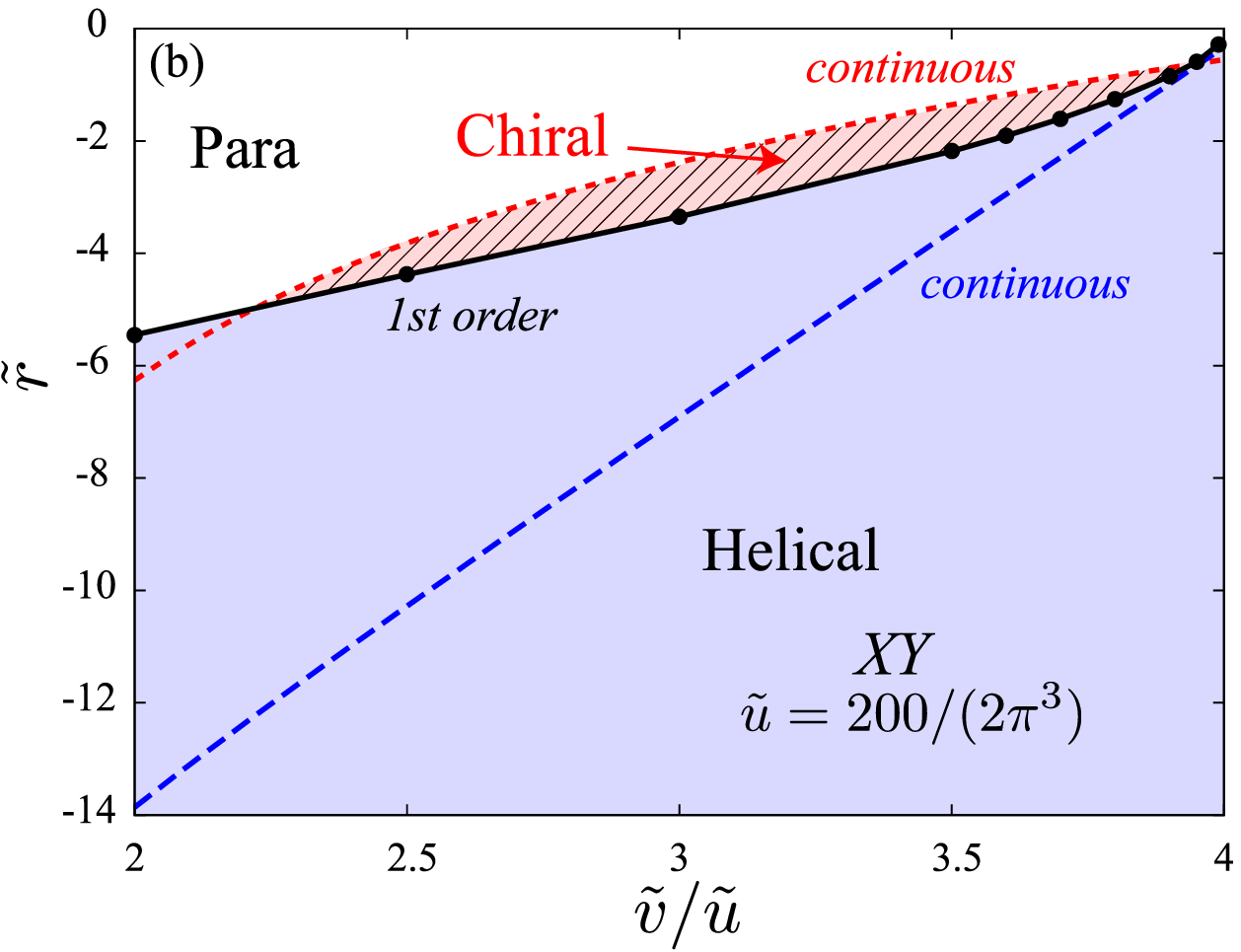}
 \caption{(Color online) Phase diagram of the {\it XY\/} ($n=2$) chiral GL model in the 
 $(\tilde{r},\tilde{v}/\tilde{u})$ plane for
  $\tilde{u}=200/(2\pi^3)$. The right figure is an enlarged view of the
  vicinity of the chiral phase. The red hatched (blue filled) area represents the chiral (helical) phase. The dotted and dashed curves represent
  continuous transition lines which would occur if the possibility of a
  first-order transition would be neglected in the analysis. The chiral phase predicted   by Onoda {\it et al} occupies the region between these two
  curves\cite{Onoda}. The solid curves represents the first-order transition line from
  the paramagnetic phase (or the chiral phase) to the helical phase. As
  in the Heisenberg case, the stability range of the chiral phase is
  largely reduced compared with that reported by Onoda {\it et al}.}
\label{XY_phase}
\end{figure*}

In the same manner as in the Heisenberg case, we find 
that a first-order transition to the helical magnetic phase occurs
and this first-order transition reduces the
stability range of the chiral phase. However, even in case of the {\it XY} spin, the
chiral phase still persists for a certain parameter range. In Fig.\ref{XY_phase} we show a typical phase diagram of the {\it XY} case. One can see
qualitatively similar structure to the one of the Heisenberg spins
(see also Fig. \ref{fig_phase}). 

\end{document}